\documentclass[11pt]{article}

\usepackage{amsfonts}
\usepackage{amssymb}
\usepackage{amsthm,amsgen}
\usepackage{graphicx}
\usepackage[mathscr]{eucal}
\usepackage{slashed}

\newcounter{subequation}
	\newenvironment{subequation}%
	{\addtocounter{equation}{-1}%
	\stepcounter{subequation}%
	\begin{equation}}%
	{\end{equation}%
}

\newcommand{\ts}{\textstyle}
\newcommand{\sgn}{\mbox{sgn}}
\newcommand{\bA}{\mathbf{A}}

\newcommand{\ba}{\mathbf{a}}

\newcommand{\bE}{\mathbf{E}}
\newcommand{\bF}{\mathbf{F}}
\newcommand{\vf}{\mathbf{f}}
\newcommand{\bG}{\mathbf{G}}

\newcommand{\bh}{\mathbf{h}}

\newcommand{\bK}{\mathbf{K}}
\newcommand{\bL}{\mathbf{L}}
\newcommand{\bM}{\mathbf{M}}

\newcommand{\bP}{\mathbf{P}}
\newcommand{\bp}{\mathbf{p}}

\newcommand{\bq}{\mathbf{q}}
\newcommand{\bR}{\mathbf{R}}

\newcommand{\bT}{\mathbf{T}}
\newcommand{\bU}{\mathbf{U}}
\newcommand{\bV}{\mathbf{V}}
\newcommand{\bv}{\mathbf{v}}

\newcommand{\bX}{\mathbf{X}}
\newcommand{\bY}{\mathbf{Y}}
\newcommand{\bZ}{\mathbf{Z}}
\newcommand{\bg}{\mathbf{g}}

\newcommand{\fe}{\mathfrak{e}}
\newcommand{\varmu}{\mathfrak{m}}
\newcommand{\fm}{\mathfrak{m}}

\newcommand{\beq}{\begin{equation}}
\newcommand{\eeq}{\end{equation}}
\newcommand{\bseq}{\begin{subequation}}
\newcommand{\eseq}{\end{subequation}}
\newcommand{\refeq}[1]{(\ref{#1})}

\newcommand{\Ric}{\mathbf{Ric}}
\newcommand{\tr}{\mathrm{tr}\,}
\newcommand{\gad}{\stackrel{\gamma}{\nabla}}

\newcommand{\fE}{\mathfrak{E}}
\newcommand{\fD}{\mathfrak{D}}
\newcommand{\fB}{\mathfrak{B}}
\newcommand{\fH}{\mathfrak{H}}
\newcommand{\p}{\partial}

\newcommand{\e}{\mathbf{e}}

\newcommand{\D}{\mathcal{D}}

\newcommand{\cK}{{\mathcal K}}

\newcommand{\cL}{{\mathcal L}}

\newcommand{\cQ}{{\mathcal Q}}
\newcommand{\cS}{{\mathcal S}}

\newcommand{\mM}{\mathcal{M}}
\newcommand{\M}{\mathcal{M}}
\newcommand{\sS}{\mathcal{S}}

\newcommand{\Rset}{{\mathbb R}}
\newcommand{\Sset}{{\mathbb S}}

\newcommand{\ze}{\zeta}

\newcommand{\la}{\lambda}

\newcommand{\de}{\delta}

\newcommand{\al}{\alpha}
\newcommand{\be}{\beta}
\newcommand{\ga}{\gamma}

\newcommand{\ep}{\epsilon}

\newcommand{\ka}{\kappa}

\newcommand{\nab}{\nabla}
\newcommand{\half}{\frac{1}{2}}

\newcommand{\bna}{\begin{eqnarray}}
\newcommand{\ena}{\end{eqnarray}}
\newcommand{\bea}{\begin{eqnarray*}}
\newcommand{\eea}{\end{eqnarray*}}
\newcommand{\ben}{\begin{enumerate}}
\newcommand{\een}{\end{enumerate}}
\newcommand{\bi}{\begin{itemize}}
\newcommand{\ei}{\end{itemize}}
\newcommand{\A}{{\mathcal A}}

\newcommand{\Hc}{\mathcal{H}}
\newcommand{\J}{\mathcal{J}}

\newcommand{\RR}{{\mathbb R}}

\newcommand{\BS}{{\mathbb S}}

\newcommand{\mb}[1]{\mathbf{#1}}

\newtheorem{thm}{THEOREM}[section]
\newtheorem*{thm*}{THEOREM}

\newtheorem{lem}[thm]{Lemma}
\newtheorem{prop}[thm]{Proposition}
\theoremstyle{definition}

\theoremstyle{remark}
\newtheorem{rem}[thm]{Remark}

\begin{document}

\title{On the Static  Spacetime of  a Single Point Charge}
\author{A. Shadi Tahvildar-Zadeh\footnote{Department of Mathematics, Rutgers, The State University of New Jersey, 110 Frelinghuysen Rd., Piscataway, NJ 08854}}

\date{orig.: Dec. 6, 2010, rev.: Feb. 7, 2011}

\maketitle

\begin{abstract}
Among all electromagnetic theories which (a) are derivable from a Lagrangian, (b) satisfy the dominant energy condition, and (c)  in the weak field limit coincide with classical linear electromagnetics, we identify a certain subclass with the property that the corresponding spherically symmetric, asymptotically flat, electrostatic spacetime metric has the mildest possible singularity at its center, namely, a conical singularity on the time axis.  The electric field moreover has  a point defect on the time axis, its total energy is finite, and is equal to the ADM mass of the spacetime.  By an appropriate scaling of the Lagrangian, one can arrange the total mass and total charge of these spacetimes to have any chosen values.  For small enough mass-to-charge ratio, these spacetimes have no horizons and no trapped null geodesics. We also prove the uniqueness of these solutions in the spherically symmetric class, and we conclude by performing a qualitative study of the geodesics and test-charge trajectories of these spacetimes.
\end{abstract}

\section{Introduction}

In this paper  we study {\em static, spherically symmetric} solutions of the Einstein-Maxwell system featuring a single spinless point charge\footnote{In a sequel to this paper we will study solutions with spin.}.  Our main results are summarized in a theorem at the end of this introduction.   The Einstein-Maxwell system of PDEs reads:
\begin{eqnarray}\label{eq:Ein}
\mathbf{R}_{\mu\nu} - {\ts\frac{1}{2}}  R \bg_{\mu\nu}&=&  \varkappa\mathbf{T}_{\mu\nu},\\
 d\bF & = & 0, \\
 d\bM & = & 0.\label{eq:Max}
\end{eqnarray}
It describes the geometry of a spacetime $(\M,\bg)$  endowed with an electromagnetic field.
Here $\mathbf{R}$ is the Ricci curvature tensor and $R$ the scalar curvature of the metric $\bg$ of a four-dimensional Lorentzian manifold $\mM$. Moreover,  $\bF$ is the Faraday tensor of the electromagnetic field and $\bM$ is the Maxwell tensor  corresponding to $\bF$(for which we will consider a whole family of choices), while   $\mathbf{T}$ is
the electromagnetic energy(density)-momentum (density)-stress tensor associated to $\bF$.  Finally,  $\varkappa = 8\pi G/c^4$, with $G$ being Newton's constant of universal gravitation and $c$ the ``speed of light in vacuum". (In the following we will work with units in which $\varkappa = 1$.)

Before we discuss the options of choosing the relationship between $\bM$ and $\bF$, we should recall that an open domain $\M'$ in a Lorentzian manifold $\M$ is (somewhat inappropriately) called {\em static} if it has a hypersurface-orthogonal Killing field $\bK$ whose orbits are complete and everywhere timelike in $\M'$.  Such a domain possesses a {\em time-function} $t$, i.e. a function defined on it whose gradient is timelike, and the vectorfield dual to its gradient is future-directed everywhere. It can always be chosen such that $\bK t = 1$. One can  use $t$ as a coordinate function on $\M'$, and the level sets of $t$ provide a foliation of $\M'$ into space-like leaves $\Sigma_t$. It follows that the induced metric on the leaves is independent of $t$, i.e. the {\em space} $\Sigma_t$ is static.  Furthermore, for a ``static spacetime'' to be {\em spherically symmetric} means that the rotation group $SO(3)$ acts as a continuous group of isometries on the manifold, with orbits that are spacelike 2-spheres, and its action commutes with that of the group generated by the timelike Killing field $\bK$.  For $p\in\M'$, let $A(p)$ be the area of the spherical orbit that goes through $p$, and let $r(p) = \sqrt{A(p)/4\pi}$ be the radius of a Euclidean sphere with area $A(p)$.  As long as $\bg(\nabla r,\nabla r) > 0$ one can use $r$ as a spacelike coordinate function on the manifold, and in that case the metric of the spacetime can be put in the form
\begin{equation}\label{gspher}
\bg_{\mu\nu}dx^\mu dx^\nu =  -e^\xi dt^2 + e^{-\xi} dr^2 + r^2(d\theta^2 + \sin^2\theta d\phi^2),
\end{equation}
 where $(\theta,\phi)$ are spherical coordinates on the orbit spheres, and $\xi=\xi(r)$ is a smooth function which depends on the choice of relationship between $\bF$ and $\bM$, here called an ``aether law" (for short, and for historical reasons.)

The simplest choice of the aether law is Maxwell's $\bM = -*\bF$, in which case the Einstein-Maxwell system will be called the Einstein-Maxwell-Maxwell system (EMM).
 It is well-known that the Reissner-Weyl-Nordstr\"om spacetime\footnote{Generally known as having been discovered independently by Reissner \cite{Rei16} and Nordstr\"om \cite{Nor18}, this spacetime is also a member of a whole class of electrovacuum solutions found by Weyl \cite{Wey17}.} (RWN for short) is the unique\footnote{See Section 6 for a precise statement and proof of uniqueness.} spherically symmetric, asymptotically flat solution
  of the EMM system.  For the RWN solution, one has
$$ e^\xi = 1 - \frac{2m_0}{r} +  \frac{q_0^2}{r^2},$$
where $q_0$ and $m_0$ are two real parameters. They are in fact integration constants that come from solving the radial Liouville-type equation that arises as the reduction of the EMM system under the stated symmetry assumptions. Since the spacetime is asymptotically flat, its ADM mass \cite{ADM61} is well defined, and it is seen to be equal to the parameter $m_0$. Also,  one has a formula for the Faraday tensor in the case of RWN:
$$ \bF = d\bA,\qquad \bA = \varphi(r) dt,\qquad \varphi = \frac{q_0}{r},$$
which suggests, via the divergence theorem, that $q_0$ is the total {\em charge} of the spacetime.

 As discussed in detail in \cite{HawEll73}, the causal structure of the RWN spacetime depends in a crucial way on the ratio $|q_0|/m_0$:  When this is less than one, which is referred to as the {\em subextremal case}, the metric coefficient $e^\xi$, in addition to being singular at $r=0$, has two zeros, namely at $ r_\pm = m_0 \pm \sqrt{m_0^2 - q_0^2}$.  It can be shown that $r=r_+$ is the {\em event horizon}, the boundary of the past of the spacetime's future null infinity, and therefore the spacetime has a nonempty black hole region.  It is worth noting that the causal structure of the maximal analytic extension of the subextremal RWN spacetime is quite rich and complicated, comprising an infinitude of regions, and is plagued by the breakdown of determinism due to the presence of Cauchy horizons (cf. \cite{HawEll73}.)  By contrast, the RWN metric in the {\em superextremal} regime, corresponding to $|q_0|>m_0$ has a very simple causal structure.  The metric coefficient $e^\xi$ is always positive,  $(t,r,\theta,\phi)$ is a global coordinate system for the manifold, and the only singularity present is the naked one, on the timelike axis $r=0$. The topology of the manifold is that of $\Rset^4$ minus a line.

 In view of the fact that the empirical charge-to-mass ratio of charged particles such as the electron and the proton are huge ($10^{18}$ and $10^{22}$ resp.)  many researchers have been tempted by the prospect that the superextremal RWN solution is but the simplest example of spacetimes featuring one or more point charges\footnote{See \cite{KraSteHerMac80}, \S 21.1 for references, and for a catalog of  such solutions.}.  Such a spacetime we shall refer to as a ``charged-particle spacetime."

One of the key questions that needs to be addressed in this regard is the following: According to relativity theory's $E=mc^2$, the proper mass of a  charged-particle spacetime should equal its energy, which for a static spacetime is expressed as the integral of the time-time component of ${\bT}$ over a static space-like hypersurface.
Is it then possible to attribute some, or all, of the mass of a charged-particle spacetime to the energy of the electromagnetic field that permeates it\footnote{Questions about the origin of the mass of charged particles are as old as the theory of electromagnetism itself, beginning with O. Heaviside, and continuing with Abraham, Lorentz, Poincar\'e, Mie, Einstein, Fermi, Born, Dirac, Wheeler, Feynman, Schwinger, Rohrlich and many others.}?

One difficulty with such an attribution is that in classical Maxwell-Lorentz electrodynamics\footnote{That is to say, Maxwell's equations with  ``point-particle-like" sources whose formal law of motion is that of Newton's, with the formal force being the Lorentz force. The model is ill-defined unless regularized.} the self-energy of a point charge is infinite, and that remains to be the case even when the electromagnetic Maxwell-Lorentz field is coupled to gravity via the Einstein equations.  This can be clearly seen in the case of RWN because the total electrostatic energy carried by a time-slice would be
$$  \int_0^\infty |d\varphi|^2 r^2 dr = \int_0^\infty \frac{q_0^2}{r^2}dr,$$
which is infinite\footnote{Indeed, this is the same infinity that turns up in the absence of gravity, in flat spacetime, for the self-energy of a point charge, and which led Abraham \& Lorentz, and later Mie, to look for alternatives to the point-charge  description. We also note that in the current approach to quantum electrodynamics, the corresponding energy integral to the above is still divergent, although less violently \cite{Wei39}, thus even in the absence of gravity, the problem of infinite self-energy of point charges is not solved by going over to the quantum description \cite{Fey64},\cite{Wei95}.}  unless $q_0 = 0$.  In general relativity, the principle of equivalence states that it is the total energy of a system that interacts gravitationally, i.e. unlike the Newtonian theory, there is no distinction between active and passive ``gravitational" masses. Thus, the total electrostatic energy will always make a contribution to the ADM mass of the spacetime, which in the case of the superextremal Reissner-Nordstr\"om is clearly an infinite contribution.

  On the other hand,  the ADM mass is usually interpreted as representing the total {\em energy} content of the spacetime (recall that in relativity there is no distinction between mass and energy (with $c=1$)).
Ingenious proposals have been made to explain the finiteness of the ADM mass, in particular to assume that the point charge possesses a ``bare mass" equal to $-\infty$. Such  ideas have been pursued using renormalization  techniques \cite{ADM60}, but these techniques are usually very difficult or even impossible to justify rigorously\footnote{For example, it is known that a rigorous removal of the regularization using mass renormalization $m_{\mbox{bare}} \to -\infty$ is impossible in the case of Lorentz electrodynamics, since the renormalization flow terminates at $m_{\mbox{bare}}=0$ with regularization still in place \cite{ApKi01}.  Similar difficulties exist for the renormalization of RWN \cite{Kie10}.}.

Another difficulty with taking the superextremal RWN too seriously is the presence of a strong naked singularity on the time axis.  Such  singularities are expected to be ``subject to cosmic censorship'', in the sense that they are believed to be non-generic, and therefore unstable under small perturbations. Furthermore, the presence of a strong ``eternal" singularity means that such a spacetime cannot arise as a solution of a classically-posed initial value problem.  For the RWN metric the worst part of the singular behavior at $r=0$ stems from the contribution of the charge $q_0$ to the metric coefficient, as can be seen for example from the Kretschmann scalar, which is a curvature invariant equal to the norm of the Riemann tensor:
$$\cK^2 =\bR_{abcd}\bR^{abcd} = \frac{48}{r^6} \left(m_0^2 - \frac{2m_0q_0^2}{r} + \frac{7q_0^4}{6r^2}\right).$$
Clearly, $\cK\sim r^{-4} \nearrow \infty$ as $r \searrow 0$.

To overcome the first of these two divergence problems while retaining the concept of a point charge, Max Born \cite{Bor33} proposed to make  the Maxwell equations nonlinear\footnote{Born picked up on the program  initiated by Mie \cite{Mie13}, who however did not want point charges.  See \cite{Kie04,Kie11} for an excellent account of the development of these ideas, and for their author's  key contribution to this program.}. This is done by choosing a Lagrangian density $\bL$ for the electromagnetic action $\sS[\bA] = \int_\mM \bL(d\bA)$ in such a way that, in addition to fulfilling the basic requirement of generating a Lorentz- and Weyl- (gauge) invariant theory,  it coincides in the weak field limit with the Lagrangian of the Maxwell-Maxwell system\footnote{which seems to have been discovered by Schwarzschild \cite{Sch03}.},
$$\bL_0 = -\frac{1}{8\pi} \bF\wedge *\bF,$$
 while its behavior in the strong field limit is such that it leads to finite total energies. One example of such a Lagrangian density is the well-known one-parameter family proposed by Born and Infeld \cite{BorInf33}:
$$\bL_\beta = *\frac{1}{4\pi\beta^4}[ 1 - \sqrt{ 1 - \beta^4 *(\bF\wedge *\bF) - \beta^8 (*(\bF\wedge \bF))^2 } ],$$
for $\beta > 0$ (in the notation of \cite{Kie04}), which even leads to finite limits of the field strengths at the location of the charge.

What is perhaps less common knowledge --though it should be equally well-known-- is that a nonlinear aether law also has the power to reduce the strength of the spacetime singularity that is present when the electromagnetic field is coupled to gravity\footnote{Even though, as we will prove in this paper, it cannot quite eliminate the singularity.}. This  phenomenon was first noticed by Hoffmann \cite{Hof35b}, who initially claimed that a solution of the Einstein-Maxwell-Born-Infeld system that he had found, was free of all singularities\footnote{Hoffmann's  enthusiasm for nonlinear electrodynamics was not dampened even after it was pointed out to him by Einstein and Rosen (See \cite{InfHof37}, fn.~15), that a mild singularity remains at the center of his spacetime.}.  In the years since the publication of Hoffmann's paper  there have been several attempts at finding static electrovacuum spacetimes which are free of all singularities \cite{InfHof37,Rao37,PelTor69,Dem86,AyoGar98,Bro01,Dym04,Cir05JMP}, either in the fields or in the metric, while at the same time various ``no-go" results have been announced \cite{BroMelShiSta79,Bro00}, that seem to show that such solutions cannot exist. One of the goals of this paper is to take steps towards dispelling the confusion that seems to persist about this subject.

Our approach here is to characterize aether laws that not only feature finite self-energies for point charges, but also  lead to {\em electrostatic}\footnote{See Section 3 for a precise definition.} spacetimes with the {\em mildest} form of singularity possible\footnote{In fact,  we will see that the remaining mild singularity in the field and the metric is of the point-defect type.  There are indications that this kind of singularity may be ``just right'' for the Hamilton-Jacobi equations to provide a law of motion for those defects \cite{Kie04,Kie11}. In this way it may turn out then  that the ``underwater stone" of nonlinear electrodynamics (in the words of \cite{Bro00}) is a gem after all!}.
Thus we will initially allow the aether law to have the most general form possible, and let the above requirements, of finiteness of the self-energy and mildness of the singularity, as well as other criteria which may arise during the course of analyzing the corresponding solutions, dictate the final form that it should have.
In particular we will prove the following:

\begin{thm*} (Informal version)
For any aether law which
{\bf(a)} agrees with that of Maxwell in the weak field limit, {\bf(b)} is derivable from a  Lagrangian, with an  energy tensor that satisfies the Dominant Energy Condition, and for which {\bf(c)} the corresponding Hamiltonian satisfies certain growth conditions (to be made precise below), the following  hold:
\vspace{-5pt}
\begin{itemize}
\item There exists a unique electrostatic, spherically symmetric, asymptotically flat solution  of the Einstein-Maxwell system \refeq{eq:Ein}-\refeq{eq:Max} with that aether law, the maximal analytical extension of which is homeomorphic to $\Rset^4$ minus a line. It has a conical singularity on the time axis\footnote{That is to say, the limit as the radius goes to zero of the ratio of the circumference to the radius of a small spacelike circle going around the axis, exists but is not equal to $2\pi$.}, which is  the mildest possible singularity for any spherically symmetric electrostatic spacetime whose aether law satisfies {\bf (a)} and {\bf(b)}.  No other singularities are present in the spacetime.
    \vspace{-5pt}
    \item A generalization of Birkhoff's Theorem shows that this solution is unique in the spherically symmetric class.
    \vspace{-5pt}
\item The electrostatic potential is finite on the axis of symmetry, which can be identified with the world line of a  point charge. The electric field has  a {\em point defect} at the location of the charge.  The field has finite total electrostatic energy, which is equal to the ADM mass of the spacetime.  The mass of the point charge is thus entirely of electromagnetic origin, i.e. it has no bare mass.
    \vspace{-5pt}
\item  By scaling the Hamiltonian appropriately, one can arrange the total mass and the total charge of this solution to have any chosen values. The mass-to-charge ratio of the spacetime enters  as a natural small parameter, measuring the departure from the Minkowski spacetime.
    \vspace{-5pt}
\item For small enough mass-to-charge ratio, there are no horizons of any kind  and no trapped null geodesics in this spacetime.
    \vspace{-5pt}
\item The analysis of geodesics and test-charge trajectories shows that the naked singularity at the center of this spacetime is gravitationally attractive (unlike the case of superextremal RWN).
\end{itemize}
\end{thm*}

The rest of this paper is organized as follows: In section 2 we introduce the Lagrangian formulation of electromagnetics. Section 3 gives the equations satisfied by an electrostatic solution of \refeq{eq:Ein}-\refeq{eq:Max}, with an arbitrary aether law.  In Section 4 we assume spherical symmetry as well, and obtain the general solution to the equations. Section 5 is devoted to the study of the singularities of this solution. In Section 6 we state and prove a Birkhoff-type uniqueness result for these spacetimes.  In Section 7 we give the precise version of our main result, and in Section 8 we carry out the qualitative analysis of geodesics and test-charge orbits.

\section{Nonlinear Electrodynamics}
Let $(\M,\bg)$ be a 4-dimensional Lorentzian manifold.  Let  $\bigwedge^p(\M)$ denote the bundle of $p$-forms on $\M$.  By an electromagnetic Lagrangian (density) we mean a mapping $\bL_{em}$ defined on  sections of the vector bundle $\bigwedge^1(\M)\times_\M\bigwedge^2(\M)$ which takes its values in $\bigwedge^4(\M)$ (see \cite{Chr99} for details.)  Thus if $\ba$ is a 1-form on $\M$ and $\vf$ a 2-form, then $\bL_{em}(\ba,\vf)$ is a 4-form on $\M$.  The electromagnetic action is by definition
$$\cS[\ba;\D] := \int_\D\bL_{em}(\ba,d\ba),$$
where $\D$ is a domain in $\M$.  A critical point of $\cS$ with respect to variations of $\ba$ that are compactly supported in $\D$ is called an electromagnetic potential $\bA$ in $\D$,
$$ \left.\frac{\delta \cS}{\delta \ba}\right|_{\ba=\bA} = 0,$$
and the exterior derivative of it is the  electromagnetic Faraday tensor $\bF = d\bA$.
The Maxwell tensor $\bM$ is by definition
\begin{equation}\label{def:M}\bM = \left.\frac{\partial \bL_{em}}{\partial \vf}\right|_{\ba=\bA,\vf=\bF} \end{equation}
in the sense of evaluation, i.e. the object on the right, when evaluated on a variation $\dot{\vf} \in T_p(\bigwedge^2(\M)) = \bigwedge^2(T_p\M)$ as a four-form is equal to $\bM\wedge \dot{\vf}$.  The source-free Maxwell equations are the Euler-Lagrange equations for stationary points of the electromagnetic action $\cS$, and are equivalent to the system
$$d\bF = 0,\qquad d\bM = 0.$$

It can be shown \cite{Chr99} that the only Lorentz-invariant gauge-invariant source-free electromagnetic Lagrangians possible are those of the form \begin{equation}\label{lem}
\bL_{em}(\ba,\vf) = -\ell(x(\vf),y(\vf))\epsilon[\bg],
\end{equation}
 where $\epsilon[\bg]$ is the volume form on $\M$ induced by the metric $\bg$, $\ell$ is a real-valued function of two variables,  and $x$ and $y$ are the electromagnetic invariants
$$
	x(\vf) := -\frac{1}{2} *(\vf\wedge *\vf) = \frac{1}{4} \vf_{\mu\nu}\vf^{\mu\nu},\qquad
	y(\vf) := \frac{1}{2} *(\vf\wedge \vf) = \frac{1}{4} \vf_{\mu\nu}*\vf^{\mu\nu}.
$$
Here $*$ is the Hodge star operator $*:\bigwedge^k(\M)\to \bigwedge^{4-k}(\M)$ with respect to the $\bg$ metric, defined by $$*\sigma_{\mu_1\dots\mu_{4-k}} = \frac{1}{k!} \sigma^{\nu_1\dots\nu_k}\epsilon[\bg]_{\nu_1\dots\nu_k \mu_1\dots\mu_{4-k}}.$$
Note that for $k$-forms on a Lorentzian 4-manifold, $** = (-1)^{k+1}$.

Furthermore, conservation of parity implies that $\ell(x,y) = \ell(x,-y)$, so that if we assume that $\ell$ is a $C^1$ function of its variables, then $$\ell_y(x,0) = 0.$$

By an {\em aether law} we simply mean a particular function $\ell(x,y)$ as the Lagrangian density function, which determines the way the electromagnetic vacuum interacts with the spacetime geometry\footnote{Traditionally, aether law referred to the relationship between tensors $\bM$ and $\bF$, similar to the constitutive relations of elastodynamics relating stresses to strains for the medium. In case of a Lagrangian theory, this is given by (\ref{def:M}), and thus the choice is that of a particular Lagrangian.}.  For example, conventional (linear) electromagnetics corresponds to the choice $\ell = x$ made first by Maxwell.

Using * one defines a dot product on $k$-forms, by $$\sigma\cdot\tau = -*(\sigma\wedge *\tau) = \frac{1}{k!}\sigma^{\mu_1\dots\mu_k}\tau_{\mu_1\dots\mu_k}.$$
 We also set $|\sigma|^2:= \sigma \cdot \sigma$ even though this is not necessarily positive.
 It follows from (\ref{lem}) that $*\!\bL_{em} = \ell$ and
$$*\bM = \frac{\partial \ell}{\partial F} = \ell_x \bF + \ell_y *\!\bF.$$

The energy tensor $\mathbf{T}$ corresponding to the Lagrangian density function $\ell$
is a symmetric 2-covariant tensor field on $\M$ defined by
$$ \mathbf{T}_{\mu\nu} = 2 \frac{\partial \ell}{\partial \bg^{\mu\nu}} - \bg_{\mu\nu} \ell,$$
which in the case of an electromagnetic Lagrangian yields
\begin{equation}
\mathbf{T}_{\mu\nu} = 2\left( \ell_x \frac{1}{2} \bF_{\mu\lambda}\bF_\nu^{\ \lambda} +\ell_y \frac{1}{2} \bF_{\mu\lambda}*\!\bF_\nu^{\ \lambda}\right) - \bg_{\mu\nu}\ell = \bF_{\mu\lambda} *\!\bM_\nu^{\ \lambda} - \bg_{\mu\nu} \ell.\label{exprT}
\end{equation}

Recall that if both of the following hold, the energy tensor $\mb{T}$  is said to satisfy the
{\sc Dominant Energy Condition:}
\begin{itemize}
\item
$ \mb{T}_{\mu\nu}\bY^\mu \bY^\nu \geq 0$ for every future-directed timelike vector $\bY$,
\item The vector $ -\mb{T}^\mu_\nu \bY^\nu$ is future directed causal when $\bY$ is a future-directed causal vector.
\end{itemize}

The first of these two is called the {\sc Weak Energy Condition.} There is also a {\sc Strong Energy Condition:} $$\left(\mb{T}_{\mu\nu} - \half \bg_{\mu\nu} \tr \mb{T}\right) \bY^\mu \bY^\nu \geq 0,$$ for all future-directed timelike $\bY$.
One can prove \cite{Ple68} that the Dominant Energy Condition is satisfied for this field theory if and only if
$$
\ell_x >0,\qquad \ell - x \ell_x -y \ell_y \geq 0.
$$
Note that under the above assumption, $\tr \bT = -4(\ell - x\ell_x -y\ell_y)\leq 0$.  We also note that it is possible for the Strong Energy Condition to be violated in nonlinear electrodynamics (even though, as we shall see, it can hold for nonlinear {\em electrostatics.})

Next we define the following one-forms, which provide a useful decomposition of the Faraday and Maxwell tensors.  Let $\bK$ be an arbitrary\footnote{In the next section we will assume that $\bK$ is a timelike Killing field for $(\M,\bg)$, but the definitions in this section are independent of that.} non-null vectorfield on $\M$, and let\footnote{Note that strictly speaking, {\em only} if $\bK$ were the unit tangent vectorfield to a timelike curve in $\M$ (the world-line of an observer)  would we be justified in calling these the (flattened) electric field, magnetic induction, electric displacement, and magnetic field, respectively.}
\begin{eqnarray}
\mathfrak{E} &:=& i_\bK \bF\label{def:al}\\
\mathfrak{B} &:=& i_\bK *\!\bF\label{def:be}\\
\mathfrak{D} &:=& i_\bK *\!\bM = \ell_x \mathfrak{E} + \ell_y \mathfrak{B}\label{def:ze}\\
\mathfrak{H} &:=& i_\bK **\!\bM = -i_\bK \bM  = -\ell_y \mathfrak{E} + \ell_x \mathfrak{B} \label{def:et},
\end{eqnarray}
where $i_\bK$ denotes the {\em interior product}  with the vectorfield $\bK$, e.g. $(i_\bK \bF)_\nu = \bK^\mu \bF_{\mu\nu}$.

For the Maxwell Lagrangian $\ell = x$ one has
$ \mathfrak{D} = \mathfrak{E}$ and $\mathfrak{B} = \mathfrak{H}$.  A general aether law will specify $\mathfrak{D}$ and $\mathfrak{B}$ as functions of $\mathfrak{E}$ and $\mathfrak{H}$, or the other way around.

Let $$X := \bg(\bK,\bK).$$   Thus $X>0$ wherever $\bK$ is spacelike, $X=0$ where $K$ is null, and $X<0$ where $\bK$ is timelike.
From the decomposition of $\bF$ in terms of $\mathfrak{E}$ and $\mathfrak{B}$ it follows that
\begin{equation}\label{defs:xy} x = \frac{|\mathfrak{E}|^2 - |\mathfrak{B}|^2}{2X} \qquad y = \frac{\mathfrak{E}\cdot \mathfrak{B}}{X}.\end{equation}
We also obtain that, by (\ref{exprT})
\begin{equation}\label{Tkk}
\mathbf{T}(\bK,\bK) = \mathfrak{E}\cdot\mathfrak{D} -X\ell,
\end{equation}
 while from (\ref{eq:Ein}) it follows that
\begin{equation}
\label{Rkk}
\mathbf{R}(\bK,\bK) = \left(\mathbf{T}(\bK,\bK)-{\textstyle\frac{X}{2}}\mbox{tr}\mathbf{T}\right) = (\mathfrak{B}\cdot\mathfrak{H}+X\ell).
\end{equation}

For $\bK$ a {\em timelike} vectorfield, we can define two electromagnetic {\em Hamiltonians} (partial Legendre transforms of the Lagrangian with respect to either $\mathfrak{E}$ or $\mathfrak{B}$)\footnote{These can also be defined for a spacelike vectorfield, but there will be some sign changes.}:
\bea H(\mathfrak{E},\mathfrak{H}) &:= &|X|^{-1}\fB\cdot\frac{\p \ell}{\p \fB} -\ell = |X|^{-1}\mathfrak{B}\cdot\mathfrak{H}-\ell,\\
\tilde{H}(\mathfrak{D},\mathfrak{B}) &:= &-|X|^{-1}\fE\cdot\frac{\p \ell}{\p\fE}+\ell = |X|^{-1} \mathfrak{E}\cdot\mathfrak{D} + \ell.
\eea
In the definition of $H$ (resp.  $\tilde{H}$) we need to think of $\mathfrak{B}$ (resp. $\mathfrak{E}$) as given by the aether law. Also, the factors of $X$ in these will disappear in the next Section, once the inner products are re-expressed in terms of a different metric on space-like slices which is {\em conformal} to the one induced by $\bg$.

We then have
$$ {\mathbf T}(\bK,\bK) = |X|\tilde{H}(\mathfrak{B},\mathfrak{D}),\qquad \mathbf{R}(\bK,\bK) = |X| H(\mathfrak{E},\mathfrak{H}).$$
The Weak, Dominant, and Strong  Energy Conditions, respectively, take the following form \cite{Tah10} in terms of the Hamiltonians
\beq\label{DECham}
 \tilde{H} \geq 0,\quad\tilde{H} \geq |H|,\quad H \geq 0.
 \eeq

It is easy to see \cite{Bia83} that $H$  is actually only a function of the three scalar invariants that one can form out of $\mathfrak{E}$ and $\mathfrak{H}$ using the metric.  More precisely,
$$ H(\mathfrak{E},\mathfrak{H}) = h(\nu, \omega, \tau),$$
where
$$
\nu := {\textstyle\frac{1}{2|X|}}(\mathfrak{E}\cdot\mathfrak{E}+\mathfrak{H}\cdot\mathfrak{H}),\qquad
\omega:= {\textstyle\frac{1}{|X|}} \mathfrak{E}\cdot \mathfrak{H},\qquad
\tau:= {\textstyle\frac{1}{2|X|}}(\mathfrak{E}\cdot\mathfrak{E}-\mathfrak{H}\cdot\mathfrak{H}),
$$
and that $h \in C^1(\Rset^3)$ is such that its gradient lies on the future unit hyperboloid in $\Rset^3$, i.e.
$$
h_{\nu}^2 - h_{\omega}^2 - h_\tau^2 = 1,\qquad h_\nu>0.$$
  Similar results hold for $\tilde{H}$, in terms of the invariants $\mu,\varpi,\sigma$ defined analogously in terms of $\mathfrak{D},\mathfrak{B}$:
 $$ \tilde{H}(\mathfrak{B},\mathfrak{D}) = \tilde{h}(\mu, \varpi, \sigma),$$
where
$$
\mu :={\textstyle \frac{1}{2|X|}}(\mathfrak{D}\cdot\mathfrak{D}+\mathfrak{B}\cdot\mathfrak{B}),\qquad
\varpi:= {\textstyle\frac{1}{|X|}} \mathfrak{D}\cdot \mathfrak{B},\qquad
\sigma := {\textstyle\frac{1}{2|X|}}(\mathfrak{D}\cdot\mathfrak{D}-\mathfrak{B}\cdot\mathfrak{B}),
$$
and likewise $\tilde{h}\in C^1(\Rset^3)$ with gradient lying on the future unit hyperboloid.

\section{Electrostatic Spacetimes}
Let $\bK$ be a timelike hypersurface-orthogonal Killing field for the spacetime $(\M,\bg)$. Let $X = \bg(\bK,\bK)<0$ and define $$\mathbf{e} = dX.$$ Let $$\bh_{\mu\nu} = \bg_{\mu\nu} - {\ts\frac{1}{X}}\bK_\mu \bK_\nu$$ denote the metric induced on the quotient $\cQ$ of $\M$ under the symmetry generated by $\bK$, and let $$\gamma = |X|\bh,$$ so that $\gamma$ is also a Riemannian metric on $\cQ$, conformal to $\bh$. Since $\bK$ is assumed to be twist-free, the quotient $\cQ$ can be identified with a spacelike hypersurface in $\M$.  Let $(x^i)$, $i=1,2,3$ be an arbitrary coordinate system on $\cQ$.  Then the line element of $\bg$ is given by
$$ ds^2 = Xdt^2 + |X|^{-1} \gamma_{ij}dx^i dx^j.
 $$
 It follows in particular that $\stackrel{\bg}{\nabla}\cdot \bV = X \gad \cdot \frac{1}{X}\bV$ for any vectorfield $\bV$ such that $\cL_\bK \bV = 0$.  The Einstein-Maxwell system in the static case reduces \cite{Tah10} to the following set of equations:
\begin{eqnarray}
\gad\cdot ({\ts\frac{1}{X}}\e) & = &{\ts\frac{-2}{X}}H(\mathfrak{E},\mathfrak{H}) \label{eq:e}\\
\gad\cdot({\ts\frac{1}{X}}\mathfrak{D}) &= & 0\label{eq:D}\\
\gad \cdot({\ts\frac{1}{X}}\mathfrak{B}) & = & 0\label{eq:B}\\
\Ric[\gamma]_{ij} -{\ts \frac{1}{2}}\gamma_{ij} R[\gamma] & = &{\ts\frac{1}{2X^2}} \e_i\e_j+{\ts\frac{1}{X} } (\mathfrak{E}_i \mathfrak{D}_j + \mathfrak{B}_i \mathfrak{H}_j)  -\gamma_{ij}\left({\ts\frac{1}{4X^2}}|\e|_\gamma^2 + {\ts\frac{1}{X} } H(\mathfrak{E},\mathfrak{H})\right),\label{eq:ricgamma}
\end{eqnarray}
where to close the system one has to remember that
\beq\label{legendre}
\mathfrak{D} = \left(\frac{\partial H}{\partial\mathfrak{E}}\right)^\flat,\qquad \mathfrak{B} = \left(\frac{\partial H}{\partial \mathfrak{H}}\right)^\flat,
\eeq
with $\flat$ being the operation of lowering indices with respect to the $\gamma$ metric.

The  equation (\ref{eq:ricgamma}) confirms that we can interpret the above as a system of Einstein equations for the manifold $\cQ$ coupled to the fields $(\e,\mathfrak{E},\mathfrak{H})$.  We further recall that Maxwell's equations for $\bF$ and $\bM$ also imply that $\mathfrak{E}$ and $\mathfrak{H}$ are exact 1-forms
$$ \mathfrak{E} = d\varphi,\qquad \mathfrak{H} = d\psi.$$
Thus the above is a system of equations for the three potentials $X,\varphi,\psi$, and the metric $\gamma$ of the quotient manifold.

We also observe that the parity conservation assumption about the Lagrangian $\ell_y(x,0)=0$ implies that, away from singularities, $\fD=0$ whenever $\fE = 0$, and likewise $\fH =0$ whenever $\fB = 0$.  This fact, together with the invariance of the equations under interchanging $\fD$ with $\fB$ and $\fE$ with $\fH$, implies that the system of equations in the {\em magnetostatic} case $\mathfrak{E} \equiv 0$ is formally the same as that in the {\em electrostatic} case $\mathfrak{H} \equiv 0$, even though (it turns out) the restrictions on the  Hamiltonian under which meaningful solutions can be obtained are different\footnote{For examples of regular magnetostatic (magnetic monopole) solutions, see \cite{Bro01}.}.

 In this paper we are confining ourselves to the electrostatic case, where $\omega  = \varpi=0$,  $\tau  = \nu$ and $\sigma = \mu$.  If we define $$\eta(\nu) = h(\nu,0,\nu),\qquad \zeta(\mu) = \tilde{h}(\mu,0,\mu), $$
then in the electrostatic case $\fD = \frac{\p H}{\p\fE} = \eta'(\nu) \fE$ and $\fE = \frac{\partial \tilde{H}}{\partial \fD} = \zeta'(\mu)\fD$.  We can moreover express $\eta$ in terms of $\zeta$.  This is because in the electrostatic case $\fB = \fH = 0$, and thus
$$H + \tilde{H} = \langle \fD , \fE\rangle_\gamma = \big\langle \fD , \frac{\partial \tilde{H}}{\partial \fD}\big\rangle_\gamma = 2\mu \frac{d\zeta}{d\mu},$$
so that we have
\beq\label{etamu} \eta(\nu) = 2 \mu \zeta'(\mu) - \zeta(\mu).
\eeq
In terms of the reduced Hamiltonian $\zeta$ the Dominant Energy
Condition (\ref{DECham}) takes the following simple form
\beq\label{DECzeta} \zeta'(\mu) > 0,\quad  \zeta -\mu\zeta'(\mu) \geq 0,\quad
\forall \mu\geq 0,\eeq
while the Strong Energy Condition reads
\beq\label{SECzeta} 2\mu \zeta'(\mu) - \zeta(\mu) \geq 0,\quad \forall \mu \geq 0.\eeq

We note that in the electrostatic case, $\nu = -x$ and $y=0$, thus $\eta(\nu) = -\ell(x,0) = -\ell(-\nu,0)$. On the other hand the function $\ze$ can be obtained  from $\ell$ via a Legendre transform: Given $\ell=\ell(x,y)$ let $f(t) := -\ell(-\half t^2, 0)$ and let $f_*(s) = \sup_{t} [st-f(t)]$ be the Legendre transform of $f$.  Then it is easy to see that $\ze(\mu) = f_*(\sqrt{2\mu})$. As an example, here are the Lagrangian density function originally proposed by Born \cite{Bor33}, and its two reduced Hamiltonians:
$$ \ell_B(x) = \sqrt{1+2x} - 1,\qquad \eta_B(\nu) = 1 - \sqrt{1-2\nu},\qquad \zeta_B(\mu) = \sqrt{1+2\mu} -1.$$

 Let
 $$ \xi := \log(-X).$$
With $\nu = \frac{1}{2}|d\varphi|^2_\gamma$ and $\fD = \eta'(\nu)d\varphi$, the electrostatic Einstein-Maxwell system then becomes
 \begin{eqnarray}
 \gad\cdot d\xi & = &2 e^{-\xi} \eta(\nu)\label{eq:xistat}\\
 \gad \cdot (e^{-\xi} \fD )& = & 0\label{eq:Dstat}\\
 \mathbf{R}[\gamma]_{ij} - \frac{1}{2}\gamma_{ij} R[\gamma] & = &\frac{1}{2} \p_i\xi\p_j\xi- e^{-\xi} \eta'(\nu)\p_i \varphi \p_j  \varphi  -\gamma_{ij}\left(\frac{1}{4}|d\xi|_\gamma^2 -  e^{-\xi} \eta(\nu)\right) \label{eq:gammastat}
 \end{eqnarray}

 \section{Spherical Symmetry}
 If we assume that the spacetime (and the electromagnetic field) in addition to being static is also spherically symmetric, there will be a further reduction in the Einstein-Maxwell system.  In particular, using the area-radius coordinate $r$ we may rewrite the metric $\gamma$ as follows
 $$
 \gamma_{ij} dx^i dx^j = dr^2 + e^\xi r^2 (d\theta^2 + \sin^2\theta d\phi^2),
 $$
 where now $\xi = \xi(r)$, and also $\varphi = \varphi(r)$,  $\fD = D(r) dr$. From (\ref{eq:Dstat}),
 $$ 0= \gad \cdot (e^{-\xi}\fD)  = \frac{1}{e^\xi r^2} \p_r( e^\xi r^2 e^{-\xi} D) = \frac{1}{e^\xi r^2}\p_r (r^2 D), $$
 and thus one obtains that $$ \fD = \frac{c}{r^2} dr,$$
where $c$ is an arbitrary constant.  On the other hand, since $ |dr|_\gamma^2 = 1$,
\begin{equation}\label{mu}
 \mu = \frac{1}{2} |\fD|_\gamma^2 = \frac{c^2}{2r^4}.
\end{equation}

From (\ref{eq:xistat}) we now obtain
$$
\frac{1}{r^2e^\xi}\frac{d}{dr}\left(e^\xi r^2 \frac{d\xi}{dr}\right) = 2 e^{-\xi} \eta(\nu).
$$
Change variable to $u = \frac{1}{r}$ to obtain
\begin{equation}\label{eq:xiu} \frac{d^2}{du^2}(e^\xi) =  c^2 \frac{\eta(\nu)}{\mu}.\end{equation}

We can now use (\ref{etamu}), integrate (\ref{eq:xiu}) twice, change the order if integration and recompute the kernel to obtain a formula for $\xi$ as a function of $r$:
\begin{equation}\label{Xsol}
e^{\xi(r)} = c'' + \frac{c'}{r} + \frac{ c^2}{r} \int_r^\infty \frac{\zeta(\mu)}{\mu} \frac{dr'}{{r'}^2},
\end{equation}
where $c,c',c''$ are arbitrary constants, and $\mu = \frac{c^2}{2 {r'}^4}$.  The requirement that the solution be asymptotically flat now forces $c'' = 1$.  On the other hand, setting $\zeta \equiv 0$ should give the Schwarzschild solution, hence $c' = -2m_0$ where $m_0$ is the mass parameter in the Schwarzschild metric.

We can also find an expression for the electrostatic potential $\varphi$ as a function of $r$.  Recall that
$$ \varphi'(r) dr = d\varphi = \fE  = \zeta'(\mu)\fD = c\zeta'(\mu) \frac{dr}{r^2},$$
and thus
$$ \varphi(r) = c\int_r^\infty \zeta'(\mu) \frac{dr'}{{r'}^2}.$$
Since in the Maxwell-Maxwell case $\zeta(\mu) = \mu$, comparison with the RWN solution shows that $c = q_0$, the charge parameter in RWN. Finally, a direct computation shows \cite{Tah10} that \refeq{eq:gammastat} is identically satisfied.

 Thus we have the following generalization of RWN to nonlinear gravitoelectrostatics:
\begin{eqnarray}
e^{\xi(r)} &=& 1 - \frac{2m_0}{r} +  \frac{2}{r}\int_r^\infty \zeta(\mu) r'^2 dr', \label{sol:xi}\\
\varphi(r) &=&  q_0 \int_r^\infty \zeta'(\mu) \frac{dr'}{{r'}^2}; \qquad \mu = \frac{q_0^2}{2 {r'}^4}.
\label{sol:phi}
\end{eqnarray}

These simple and elegant formulae seem to have first appeared in \cite{PelTor69}, with a different derivation, although special cases of it were known long before (see below.)
Note that only the reduced Hamiltonian $\zeta$ makes an appearance in them.
The prospect of generating exact solutions to the Einstein-Maxwell system with interesting and desirable properties, just by inserting a judiciously chosen $\zeta$ into these formulae has proven to be irresistible to many theoreticians. In particular  we should mention the solution found by Hoffmann \cite{Hof35} to the Einstein-Maxwell- Born-Infeld system, which corresponds to the following Hamiltonian:  \beq\label{BIham}\zeta_{BI}(\mu) = \sqrt{1+2\mu}-1.\eeq
  Another early example is the solution found by Infeld and Hoffmann \cite{InfHof37}, where they made the following choice \beq\label{IHham}\zeta_{IH}(\mu) = \log(1+\mu),\eeq
  and obtained a metric which was completely smooth and free of all singularities!
  Their work was followed up by Rao \cite{Rao37}, who attempted to find a large family of actions leading to such solutions.

   Infeld and Hoffmann however may also be the first of many investigators who have made the mistake of picking a Hamiltonian that is {\em not admissible} because it cannot arise from a Lagrangian:
One has to remember that $\zeta(\mu)$ is the electrostatic reduction of the Hamiltonian $\tilde{H}(\fB,\fD)$, which in turn is subject to the following restrictions:
\begin{enumerate}\item It must in the weak field limit agree with Maxwell's choice for the aether law.
\item It must correspond to an energy tensor that satisfies the dominant energy condition.
\item It must be the Legendre-Fenchel transform in $\fE$ of a Lagrangian density $\ell =\ell(\fE,\fB)$, i.e., it must be convex in $\fD$.
\end{enumerate}
It follows from the above that the function $\zeta$ is subject to the following requirements:
\begin{itemize}
\item[{\bf (R1)}] $\lim_{\mu\to 0} \frac{\zeta(\mu)}{\mu} = 1$.
\item[{\bf(R2)}]
 $\zeta'(\mu) >0$ and $\zeta - \mu\zeta'(\mu)  \geq 0$ for all $\mu>0$.
\item[{\bf (R3)}] $\zeta'(\mu) + 2\mu\zeta''(\mu) \geq 0$ for all $\mu>0$.
\end{itemize}
The Hamiltonian (\ref{IHham}) proposed by Infeld and Hoffmann violates the third condition above.  Therefore it cannot  arise from a single-valued Lagrangian.

The  condition  {\bf (R3)} is equivalent to insisting that for the above solution (\ref{sol:phi}), $\nu = -\frac{1}{2} (\varphi'(r))^2$ be monotone decreasing in $r$ (note that $\mu=q_0^2/(2r^4)$ is {\em always} monotone, independent of the choice of a Lagrangian.) It was shown by Bronnikov et al \cite{BroMelShiSta79} that this monotonicity requirement of $\nu$ {\em rules out} the possibility of having electrostatic solutions with a regular center (i.e. no curvature blowup at $r=0$) that are at the same time Maxwellian in the weak field limit. The same argument applies to show that the three conditions above are incompatible with having a regular center\footnote{Another possibility is of course, not to have a center at all \cite{Bro01}.  The topology of such a spacetime however, does not seem to lend itself to the point-charge concept.}.  Therefore spacetimes corresponding to (\ref{sol:xi}-\ref{sol:phi}) must have some kind of a singularity at $r=0$.  In view of this fact, fantastic claims about existence of singularity-free  point-charge metrics in nonlinear electrodynamics, which every now and then appear in the literature, should be viewed with a healthy dose of skepticism, and the ``Hamiltonian" involved should be examined carefully, for it may violate one or more of the above conditions.

We note also that {\bf (R3)} implies in particular that the Strong Energy Condition, which we mentioned before can   be violated in nonlinear electrodynamics, nevertheless holds in the electrostatic case, since $0\leq 2\mu\ze''+\ze' = (2\mu\ze' - \ze)'$, which upon integration on $[0,\mu]$ and using {\bf (R1)} implies that \refeq{SECzeta} holds. This will have important consequences, as we will see below.

Interestingly, a question that should have been addressed long ago, but wasn't, is this:  Since the above requirements rule out solutions that are everywhere regular,  what is then the {\em mildest} singular behavior allowed by them? In the next section we characterize those static spherically symmetric point-charge metrics that have the mildest form of singularity possible at their center.

\section{Singularities}
\subsection{Singularities in the metric}
We begin by calculating the curvature tensor of a static spherically symmetric metric of the form $$ds^2 = f^2 dt^2 - f^{-2} dr^2 + r^2 d\Omega^2,$$
with $f=f(r)$.  The nonzero components of the Riemann tensor are
$$
\bR_{0101} = f f''+{f'}^2,\quad \bR_{0202} = \bR_{0303} = - \bR_{1212} =- \bR_{1313} = r^{-1}f f',\quad \bR_{2323} = r^{-2}(1-f^2).
$$
The indices here refer to the rigid frame $\{\omega^{(\mu)}\}$ defined as follows:
$$
\omega^{(0)} = f dt,\quad \omega^{(1)} = f^{-1}dr,\quad \omega^{(2)} = r d\theta,\quad \omega^{(3)} = r\sin \theta d\phi.
$$
Thus the Kretschman scalar is, with $X = -f^2$,
\begin{equation}\label{eq:kretch} \mathcal{K}^2 = {X''}^2 +  r^{-2} {X'}^2 + r^{-4}(1+X)^2.\end{equation}
It is evident from (\ref{eq:kretch}) that for there to be no spacetime curvature singularity at $r=0$ it is necessary and sufficient that $1+X(r) =O_2( r^2)$ as $r\to 0$.\footnote{For a continuous function $f:\Rset^+\to \Rset$,  integer $k$ and $\alpha>0$ we say that $f = O_k(r^\alpha)$ if $\lim_{r\to 0}r^{j-\alpha}d^j f/dr^j$ exists and is finite for $j=0,\dots,k$.} More generally, the Kretschman scalar $\mathcal{K}$ will blow up like $r^{-\alpha}$ if and only if $1+X(r) = O_2(r^{2-\alpha})$.

For spherically symmetric, electrostatic spacetimes,
$$ X = -1 + \frac{2}{r} m(r),$$
where $$ m(r) := m_0 -  \int_r^\infty \zeta\left(\frac{q_0^2}{2r'^4}\right) r'^2 dr'$$
is the {\em mass function}.

Thus we see that $\mathcal{K}$ will blow up at least like $r^{-3}$ if $m(0) \ne 0$, as it is for example in the case of the ``negative mass" Schwarzschild solution, where $m_0<0$ and $q_0 = 0$.  For the superextremal RWN solution, the situation is much worse since $m(0) = -\infty$.
Since our goal here is to characterize solutions which are as mildly singular  as possible at the location of the charge, we may start by requiring $m(0) = 0$.  Now, since
$$ m(0) = m_0 - \frac{  |q_0|^{3/2}}{2^{11/4}}\int_0^\infty y^{-7/4}\zeta(y) dy = m_0 -|q_0|^{3/2}I_\zeta$$
and $m_0$ is an integration constant, it is always possible to meet this requirement as long as \beq\label{cond4}
I_\zeta:=2^{-11/4}\int_0^\infty y^{-7/4}\zeta(y) dy < \infty.\eeq  From now on we will add this to the list of requirements that the reduced Hamiltonian $\zeta$ must satisfy:
  \begin{itemize}
  \item[{\bf(R4)}] $$\int_0^\infty \mu^{-7/4}\zeta(\mu) d\mu < \infty.$$
  \end{itemize}
Note that this new restriction implies the following:  If $\zeta(\mu)$ is assumed to grow like a power $\mu^\alpha$, then we must have $\alpha < 3/4$, which of course rules out the Maxwellian case.

Having made such a choice of the integration constant $m_0$, we now observe that
$$m_{ADM} = m(\infty) = |q_0|^{3/2}I_\zeta.$$
This means that {\em the mass of this spacetime is entirely of electrical nature}.  Moreover, by an appropriate scaling of the aether law, namely $\ell_\beta(x,y):= \beta^{-4}\ell(\beta^4 x,\beta^4 y)$, which  for the reduced Hamiltonian amounts to using $$\zeta_\beta(\mu) = \beta^{-4} \zeta(\beta^4 \mu)$$ in place of $\zeta$, it is possible to ``fit" the ADM mass and total charge of this solution to  the empirical  mass and charge of any particle. This is because $I_{\zeta_\beta} = \beta^{-1} I_\zeta$, so for a given pair of numbers $(\textsc{q},\textsc{m})$ setting $q_0 = \textsc{q}$ and \beq\label{valbe}
\beta := \frac{ \textsc{q}^{3/2}I_\zeta}{\textsc{m}}
\eeq
will result in $m(\infty) = \textsc{m}$, while at the same time the scaled version of an admissible Hamiltonian function $\zeta$ will remain admissible.   In this connection it is worth mentioning that if one carries out this procedure for the Born-Infeld Lagrangian, with $m_0$ and $q_0$ set to the mass and charge of the electron, the value for the scaling parameter $\beta$ thus obtained coincides with the value originally proposed by Born \cite{Kie04}.

Once $m_0$ is chosen as above, the mass function can be rewritten as follows:
\begin{equation}\label{mass} m(r) =  \int_0^r \zeta_\beta\left(\frac{\textsc{q}^2}{2r'^4}\right) r'^2 dr' = \textsc{m} - \int_r^\infty \zeta_\beta\left(\frac{\textsc{q}}{2{r'}^4}\right) {r'}^2 dr'.\end{equation}
We can  now obtain some estimates for $m(r)$.  Recall the second part of {\bf(R2)}:
$$ \zeta(\mu) \geq \mu\zeta'(\mu).$$ Integrating this inequality on $[0,\mu]$ and using {\bf(R1)} we easily obtain $$\zeta(\mu) \leq \mu \qquad \forall \mu >0.$$
This gives the following lower bound for the mass function:
$$ m(r) \geq \textsc{m} - \frac{\textsc{q}^2}{2r},$$
which in turn gives the following bound on the metric coefficient $e^\xi$:
$$e^\xi \leq 1 - \frac{2\textsc{m}}{r} + \frac{\textsc{q}^2}{r^2}.$$
These two estimates are clearly only useful for large $r$.  In fact, assuming slightly more than {\bf(R1)}, one can turn these into large-$r$  asymptotics for $m$ and $e^\xi$.  Namely, let us assume that
\begin{itemize}\item[{\bf(R1)'}]$$\zeta(\mu) = \mu + O(\mu^{5/4}),\qquad \mbox { as } \mu \to 0.$$\end{itemize}
Substituting into the second expression  for the mass function in \refeq{mass} we obtain
 $$ m(r) = \textsc{m} - \frac{\textsc{q}^2}{2r} + O\left(\frac{1}{r^2}\right),\quad e^\xi = 1 - \frac{2\textsc{m}}{r} + \frac{\textsc{q}^2}{r^2} + O\left(\frac{1}{r^3}\right).$$
 As advertised, these asymptotics are identical to those of the RWN solution.

 We now need to establish the small-$r$ behavior of the mass function.
 Clearly, $m(r) = O_2(r^3)$ if and only if $\zeta$ goes to a constant  as $\mu \to \infty$.  On the other hand, we recall {\bf(R3)} on $\zeta$, i.e. the convexity requirement. It is equivalent to
\beq\label{convexity} (\mu^{1/2} \zeta'(\mu))' > 0.\eeq
Integrating this on an interval $[\mu_0,\mu]$ for a fixed $\mu_0>0$, we obtain
$$ \zeta(\mu) \geq C_1 +C_2 \sqrt{\mu},$$
with $C_1,C_2\ne 0$ constants depending on $\mu_0$.  Therefore the reduced Hamiltonian must grow at least like $\mu^{1/2}$ in order to satisfy this requirement. In particular, it cannot go to a constant at infinity, hence there will be curvature blowup at $r=0$ no matter what aether law is chosen, as anticipated in \cite{BroMelShiSta79}.

Assuming now that the reduced Hamiltonian grows at the slowest possible rate, namely like $\mu^{1/2}$, we obtain that there must be a {\em conical} singularity at $r=0$ where $\mathcal{K}$ blows up like $r^{-2}$, and that there are no horizons for this metric.  In order to do this  rigorously, we need to make the growth condition more precise:
\begin{itemize}
\item[{\bf(R5)}] There exists positive constants $J_\ze, K_\ze, L_\ze$ depending only on the profile $\ze$ such that $$ J_\ze \sqrt{\mu} - K_\ze \leq \ze(\mu) \leq J_\ze \sqrt{\mu},\qquad \frac{J_\ze}{2\mu^{1/2}} - \frac{L_\ze}{\mu} \leq \ze'(\mu) \leq  \frac{J_\ze}{2\mu^{1/2}}.$$
    \end{itemize}
We should here pause to mention that an example of a reduced Hamiltonian satisfying all five requirements {\bf (R1-5)} is the one of Born-Infeld \refeq{BIham}.  Many other examples can easily be constructed by considering smooth, monotone increasing, concave functions of $\mu$ that behave like $\mu$ for small $\mu$ and like $c\sqrt{\mu}$ for large $\mu$.

Assuming {\bf(R5)}, from \refeq{mass} we have,
\beq\label{est:m} m(r) \leq \frac{A}{2}\ep^2 r,\qquad A :=\frac{\sqrt{2} J_\zeta}{I_\zeta^2},\quad \ep:= \frac{\textsc{m}}{|\textsc{q}|}.\eeq
This right away implies that there will be no horizons as long as $\ep$ is small enough, since
$$-X = 1-\frac{2m(r)}{r} \geq 1-A\ep^2>0.$$
Combining {\bf(R5)} with our previously obtained bounds for the reduced Hamiltonian,
\beq\label{zetabounds}\begin{array}{rcl} \max\{0, J_\zeta \sqrt{\mu} - K_\zeta\}  \leq &\zeta(\mu) &\leq \min\{\mu,J_\zeta\sqrt{\mu}\},\\[12pt]
\max\{ 0, \frac{J_\zeta}{2\mu^{1/2}} - \frac{L_\ze}{\mu}\} \leq & \ze'(\mu) & \leq \min\{ 1, \frac{J_\zeta}{2\mu^{1/2}}\}.\end{array}\eeq
We can now use this to obtain a lower bound for the mass function that does not degenerate near $r=0$, namely
\beq\label{lowerboundm} m(r) \geq \frac{A\ep^2}{2} r - \frac{K_\zeta}{3\beta^4} r^3,\eeq
as well as the following small $r$ asymptotics for $m$:
$$m(r) = \frac{A\epsilon^2}{2}r - \frac{B^2\ep^6}{2\textsc{m}^2} r^3 + o_2(r^3),\qquad B^2:=\frac{2K_\ze}{3 I_\ze^4}.$$
Consequently, $$-X= e^\xi = a^2+ b^2r^2 + o_2(r^2)$$
 with $a=\sqrt{1-A\ep^2}$ and $b=\frac{B}{\textsc{m}}\ep^3$.  This in turn implies that there is a conical singularity at $r=0$, because the line element of the spacetime metric is $Xdt^2 - X^{-1} dr^2 + r^2 (d\theta^2 + \sin^2\theta d\phi^2)$, and the coefficient of $dr^2$ at $r=0$ is $\frac{1}{a}$, which is greater than one.  In fact, introducing standard Cartesian space coordinates $(x^1,x^2,x^3)$ near $r=0$, with $r = (\sum_{i=1}^3 (x^i)^2)^{1/2}$ we see that the line element in these coordinates is
 $$X dt^2 + \sum_{i,j}\left[\left(\frac{1}{|X|} -1\right)\frac{x^i x^j}{r^2} + \delta_{ij}\right] dx^i dx^j,$$
 thus the metric has no continuous extension at $r=0$ unless $a=1$.  We therefore need to take the axis $r=0$ out of the spacetime manifold $\M$, which gives it the topology of $\RR^4$ minus a line.

We also note that, given a profile $\zeta$, the deficit angle of the conical singularity is proportional to $\ep=\textsc{m}/|\textsc{q}|$.  This will be quite small precisely when the empirical charge-to-mass ratio of the particle to which this solution is being fitted is large, which happens to be the case for the electron and the proton, etc.  This means that in the study of these metrics it is permissible to consider the small $\epsilon$ regime.

The last observation we would like to make about the metric before moving on to discussing the electric field, is that the metric coefficient $e^\xi = -X$ is monotone increasing.  The easiest way to see this is from \refeq{eq:xiu}.  Recall that the assumption {\bf(R3)} implies that the Strong Energy Condition is satisfied, and thus $\eta(\nu)\geq 0$, so that $e^\xi$ is convex as a function of $u=\frac{1}{r}$.  We have already established that $a^2\leq e^\xi \leq 1$, $e^\xi|_{u=0} = 1$, $\lim_{u\to\infty} e^\xi = a^2<1$.  Thus $e^\xi$ cannot have a local maximum or a local minimum at a finite $u$, and must therefore be monotone decreasing in $u$, hence monotone increasing in $r$. We will see that this fact, which is in great contrast to the behavior of the same metric coefficient in the RWN spacetimes, has important consequences, in particular for the trajectories of test particles.

\subsection{Singularities in the electric field}
In the spherically symmetric, electrostatic case, the only nonzero component of the Faraday tensor is $\bF_{rt} = \varphi'(r)$ and the one-form $\fE = i_\bK \bF = d\varphi$.  We have
$$ \varphi(r) = \textsc{q}\int_r^\infty \ze'\left(\frac{\textsc{q}^2\be^4}{2r'^4}\right)\frac{dr'}{r'^2}.$$
One easily computes that
$$ \sgn(\textsc{q})\varphi(0) = \frac{3}{2} \ep, \qquad \sgn(\textsc{q})\varphi'(0) = -\frac{A}{2\textsc{m}} \ep^3.$$
Moreover, using \refeq{zetabounds}, we have
\bea\max\left\{0, \frac{3\ep}{2} - \frac{A\ep^3}{2\textsc{m}} r \right\} \leq &\sgn(\textsc{q})\varphi(r)&\leq \min\left\{ \frac{|\textsc{q}|}{r}, \frac{3\ep}{2}- \frac{A\ep^3}{2\textsc{m}} r+ \frac{C\ep^7}{3\textsc{m}^3}r^3 \right\},\\
 \max\left\{ -\frac{|\textsc{q}|}{r^2}, -\frac{A\ep^3}{2\textsc{m}}\right\}\leq &\sgn(\textsc{q})\varphi'(r)& \leq \min\left\{ 0, -\frac{A\ep^3}{2\textsc{m}} + \frac{C\ep^7}{\textsc{m}^3}r^2\right\},
\eea
with $C:= 2L_\ze/I_\ze^4$.
In particular, $\varphi$ is monotone decreasing, and is asymptotic to the Coulomb potential as $r\to \infty$.  Moreover, $\varphi'(0) \ne 0$, and thus $\fE$ becomes undefined at $r=0$.  More precisely, since $\fE = \varphi'(r) dr$ and $dr$ is a unit covector whose direction is undefined at $r=0$, it follows that the covectorfield $\fE$ is of finite magnitude and undefined direction, i.e. has  a {\em point defect}  at $r=0$.  $\fE$ is otherwise smooth.

We now check that the total electrostatic energy is finite, and in fact is equal to the ADM mass:  By virtue of \refeq{eq:Ein}, the energy tensor $\mathbf{T}$ is divergence free, i.e. $\nabla^\mu \mathbf{T}_{\mu\nu} = 0$.  Let $\bP_\mu := -\mathbf{T}_{\mu\nu}\bK^\nu$ where $\bK=\p_t$ is the timelike Killing field of $(\M,\bg)$.  It follows that $\delta \bP = *d*\bP = 0$ and thus by the divergence theorem $*\bP$ is a conserved current, i.e. $\int_{\Sigma_t} *\bP$ is independent of $t$.  With coordinates $(t,r,\theta,\phi)$ as before, $\bK=(1,0,0,0)$ and we calculate that the only nonzero component of $*\bP$ is
$$
(*\bP)_{123} = -\bg^{00}\bP_0 \sqrt{-\det \bg} ={\textstyle\frac{-1}{X}} \bT(\bK,\bK) \sqrt{-\det \bg} = \tilde{H}(\fB,\fD) \sqrt{-\det \bg} = \zeta(\mu) r^2 \sin\theta.$$
Thus, defining the total electromagnetic energy carried by the slice $\Sigma_t$  to be
$$ \textsc{E} := \frac{1}{4\pi} \int_{\Sigma_t} *\bP,$$
we see that for the particle-spacetimes under study
$$\textsc{E} = \frac{1}{4\pi} \int_0^{2\pi}\int_0^\pi\int_0^\infty (*\bP)_{123}\ dr d\theta d\phi = \int_0^\infty \zeta(\mu) r^2 dr = \textsc{m},$$
as promised.

\section{Uniqueness in the Spherically Symmetric Class}
Birkhoff's celebrated theorem \cite{Bir23} states that any spherically symmetric solution of vacuum Einstein's equations $\mathbf{R}_{\mu\nu} = 0$ is locally isometric to a region in Schwarzschild spacetime\footnote{This result is often paraphrased inaccurately as ``any spherically symmetric vacuum solution of Einstein's equations must be {\em static},'' which makes it no longer true in general \cite{EhlKra06}.  The theorem was discovered first by J. T. Jebsen \cite{Jeb21}, whose proof appeared two years before Birkhoff's.  Jebsen's proof however, contains an error \cite{EhlKra06}.  There is a more general result, by J. Eiesland \cite{Eie25}, on necessary and sufficient conditions for the existence of an {\em extra} Killing field for spherically symmetric (not necessarily vacuum) spacetimes, a preliminary version of which was also announced \cite{Eie21} before Birkhoff's book but the final paper appeared two years after it. Birkhoff's theorem is a corollary of Eiesland's result.}.  This was generalized to the electrovacuum case by Hoffmann \cite{Hof32}, thereby proving the uniqueness of the RWN solution in the spherically symmetric class.  Many other extensions and generalizations have followed since, see \cite{SchWit10} and references therein.  Here we prove that the charged-particle spacetime whose existence we have established in the above, enjoys the same uniqueness property.

To begin, we recall the definition of spherical symmetry: A Lorentzian manifold $(\M,\bg)$ is {\em locally spherically symmetric} if every point in $\M$ has an open neighborhood $V$ such that
\begin{enumerate}
\item  There is a linearly independent set of vector fields $\{\bU_i\}_{i=1}^3$ on $V$ which generate a faithful representation of the Lie algebra of the rotation group $SO(3)$, and whose orbits are two-dimensional and spacelike. (Note that the orbits do not need to be contained in $V$).
\item $\cL_{\bU_i} g = 0$ in $V$ for $i=1,2,3$, i.e. $\bU_i$ are Killing vector fields for the metric.
\end{enumerate}
Next we recall the following (see \cite{Eie25} for a proof):
 \begin{prop}\label{prop:sphsym}
  The necessary and sufficient condition for local spherical symmetry of a smooth Lorentzian manifold $(\M,\bg)$ is that every point in $\M$ has a neighborhood in which there exists a system of local coordinates $(t,r,\theta,\phi)$  such that the line element of the metric in those coordinates has the form \beq\label{sphsymmetric}ds^2= - A^2(t,r) dt^2 + B^2(t,r) dr^2 + C^2(t,r) (d\theta^2 + \sin^2\theta d\phi^2),\eeq
with $A,B,C$ smooth functions of $(t,r)$\footnote{Jebsen's error was that he had assumed $C(t,r)\equiv r$.  This can only be achieved if $\nabla C$ is spacelike, thus for a complete proof one had to consider three other cases as well, of $\nabla C$ being timelike,  null, or  zero.}.
\end{prop}
 In this case, the three Killing fields generating  spherical symmetry are
$$\bU_1 = \sin\phi\p_\theta + \cot\theta\cos\phi\p_\phi,\quad \bU_2 = \cos\phi\p_\theta - \cot\theta\sin\phi\p_\phi,\quad \bU_3 = \p_\phi,$$
and we can check that these vectorfields satisfy the commutation relations of the Lie algebra of the rotation group: $$[\bU_i,\bU_j] = \ep_{ij}^k \bU_k,$$ where $\ep_{ijk}$ is the completely antisymmetric 3-symbol.  We also note that $[\bU_i,\bZ] = 0$ if $\bZ$ is either $\p_t$ or $\p_r$.  Moreover, we can express $\p_\theta$ and $\p_\phi$ in terms of $(\bU_1,\bU_2)$:
\beq\label{thetaphi} \p_\theta = \sin\phi \bU_1 + \cos\phi \bU_2,\qquad \p_\phi = \tan\theta(\cos\phi \bU_1 - \sin\phi \bU_2).\eeq

Suppose now that $\bF = \bF_{\mu\nu}dx^\mu \wedge dx^\nu$ is a spherically symmetric 2-form defined in a neighborhood $V$ of a spherically symmetric manifold $(\M,\bg)$, i.e. $\cL_{\bU_i}\bF = 0$ for $i=1,2,3$.  Let $\bX,\bY,\bZ$ be any three vectorfields on $\M$.  For any 2-form $\bF$ we have the identity
$$ (\cL_\bX \bF)(\bY,\bZ) = \bX\bF(\bY,\bZ) - \bF([\bX,\bY],\bZ) - \bF(\bY,[\bX,\bZ]).$$
Thus, for a spherically symmetric $\bF$, we obtain the following
\begin{enumerate}
\item $\bU_i \bF(\p_t,\p_r) = 0$, $i=1,2,3$.
\item $\bU_i \bF(\bU_j,\bZ) = \ep_{ij}^k \bF(\bU_k,\bZ)$ for $i,j=1,2,3$ and $\bZ \in \{ \p_t,\p_r\}$.
\item $\bU_i F(\bU_j,\bU_k) = 0$ for $i,j,k$ distinct, and $\bU_i\bF(\bU_i,\bU_j) = \ep_{ij}^k \bF(\bU_i,\bU_k)$ for $i=1,2,3$.
\end{enumerate}
Thus, from item (1), $\p_\theta \bF(\p_t,\p_r) = \p_\phi \bF(\p_t,\p_r) = 0$ which implies that
\beq\label{Ftr} \bF(\p_t,\p_r) = f(t,r).\eeq
From item (2) we have $\bU_1 F(\bU_3,\bZ) = -\bF(\bU_2,\bZ)$, $\bU_2 \bF(\bU_3,\bZ) = \bF(\bU_1,\bZ)$ and $\bU_3 \bF(\bU_3,\bZ) = 0$, giving a differential equation for $\bF(\bU_3,\bZ)$, and solving that we obtain that for each choice of $\bZ$ there is a function $a(t,r)$ such that
\beq\label{FphiZ} \bF(\p_\phi,\bZ) = a(t,r)\sin\theta.\eeq
On the other hand, item (2) also implies
$ \bU_3 \bF(\bU_1,\bZ) = \bF(\bU_2,\bZ)$, $\bU_3 \bF(\bU_2,\bZ) = - \bF(\bU_1,\bZ)$, which gives a differential system for $\bF(\bU_i,\bZ)$, for $i=1,2$.  Solving it, making use of \refeq{thetaphi}and \refeq{FphiZ} one obtains that for each choice of $Z$ there is a function $b=b(t,r)$ such that
\beq\label{FthetaZ}\bF(\p_\theta,\bZ) = b(t,r).\eeq
Finally, in a similar manner item (3) gives rise to a differential system for $\bF(\bU_3,\bU_i)$, $i=1,2$, which upon solving yields that there is a function $c(t,r)$ such that
$$\bF(\p_\theta,\p_\phi) = c(t,r)\sin\theta.$$
We have thus shown that a spherically symmetric 2-form $\bF$ on a spherically symmetric manifold, with coordinates $(t,r,\theta,\phi)$ as above, must have  the following form
\beq \label{sphsymF} \begin{array}{ccc} \bF_{tr} = f(t,r) & \bF_{t\theta} = b_1(t,r) & \bF_{r,\theta} = b_2(t,r)\\
 \bF_{t\phi} = a_1(t,r)\sin\theta &  \bF_{r\phi} = a_2(t,r)\sin\theta & \bF_{\theta\phi} = c(t,r)\sin\theta.\end{array}\eeq

Now, assume that $\bF$ is a {\em closed} 2-form: $d\bF = 0$.  Using the identity
$\cL_\bX \bF = i_\bX d\bF + d i_\bX \bF$ valid for any vectorfield $\bX$ and any tensor $\bF$ we obtain that for a spherically symmetric closed 2-Form $\bF$,
 $$ d i_{\bU_j} \bF = 0, \quad j=1,2,3.$$
Writing the resulting differential equations for the components of $\bF$ as found in \refeq{sphsymF} one then sees that the only solution is
$$ a_1 = a_2 = b_1 = b_2 = 0,\qquad c(t,r) = c,$$
with $c$ an arbitrary constant.  Thus we have proven the following Lemma (stated without proof in \cite{Hof32}):

\begin{lem}\label{lem:sphsym} On a spherically symmetric manifold $\M$ with coordinates $(t,r,\theta,\phi)$ the only nonzero components of a spherically symmetric Faraday tensor $\bF$ are
\beq \label{Fcomp} \bF_{tr} = f(t,r),\qquad \bF_{\theta\phi} = c\sin\theta,\eeq
where $f$ is an arbitrary function and $c$ an arbitrary constant.
\end{lem}
Note that the constant $c$ here is the total {\em magnetic charge}:
$$c = \frac{-1}{4\pi}\int_S \bF,$$
where $S$ is any spacelike 2-sphere in $\M$ that contains the origin.

We can now compute the energy tensor $\mathbf{T}$.  Recall that the line element of the metric $\bg$ has the form \refeq{sphsymmetric}. The nonzero components of the dual tensor $*\bF$ are therefore
$$ *\bF_{tr} = \frac{AB}{C^2}c,\qquad *\bF_{\theta\phi} = \frac{-C^2}{AB}f(t,r) \sin\theta ,$$
 and the dual  to the Maxwell tensor $*\bM = \ell_x\bF + \ell_y *\bF$ is computed to have components
 \beq\label{starMcomp} *\bM_{tr} = \ell_x f + c \ell_y \frac{AB}{C^2},\qquad *\bM_{\theta\phi} = \left(c \ell_x - \ell_y \frac{C^2}{AB} f\right)\sin\theta.\eeq

 We now quote the main theorem in \cite{Eie25} (slightly reworded to match our notation):
 \begin{thm} (Eiesland, 1925) The necessary and sufficient conditions that a  locally spherically symmetric Lorentzian manifold with line element
 $$ ds^2 = -A^2 dt^2 + B^2 dr^2 + C^2 (d\theta^2 + \sin^2\theta d\phi^2),$$
 $A,B$ and $C$ being arbitrary functions of $t$ and $r$, and $C$ not a constant, shall admit a one-parameter group of isometries generated by a vectorfield $\bK = k_0(t,r)\p_t + k_1(t,r)\p_r$ are
 \bna A^2 \mathbf{G}_r^t  & =  & \Psi' \p_r C \p_t C,\\
 AB(\mathbf{G}_r^r - \mathbf{G}_t^t) & = & \Psi' C\left[ A^2 (\p_rC)^2 + B^2 (\p_t C)^2\right],
 \ena
 where $\mathbf{G}_\mu^\nu = \mathbf{R}_\mu^\nu - \half R \de_\mu^\nu$ is the Einstein tensor and $\Psi=\Psi(C)$ is an arbitrary function of $C$, or a constant. If the above conditions are satisfied, then the Killing field $\bK$ is, up to a constant multiple, given by
 $$ k_0(t,r) = - \frac{1}{e^\Psi AB}\p_r C ,\qquad k_1(t,r) = \frac{1}{e^\Psi AB}\p_t C.$$
 \end{thm}

We are now in a position to state
our uniqueness result:
\begin{thm}\label{thm:uniq}
Let $(\M,\bg,\bF)$ be a locally spherically symmetric solution of the Einstein-Maxwell system \refeq{eq:Ein}-\refeq{eq:Max}. There exists a system of local coordinates $(t,r,\theta,\phi)$ on $\M$ in which the line element of the metric $\bg$ takes the form  \refeq{sphsymmetric}.  If $\nab C$ is spacelike in $\M$, then there exists a  Killing vectorfield $\bK$ which is hypersurface-orthogonal and timelike everywhere in $\M$, i.e. the solution is static as well, and in the case of no magnetic charge it is thus isomorphic to a region in the electrostatic charged-particle spacetime found in our Section 4.
\end{thm}

{\em PROOF}:  The existence of the coordinate system is guaranteed by Prop.~\ref{prop:sphsym}. By Lemma~\ref{lem:sphsym} the Faraday tensor $\bF$ in these coordinates has the form \refeq{Fcomp}.
 On the other hand, by virtue of \refeq{eq:Ein} we must have $\mathbf{G}_{\mu}^{\nu} = \mathbf{T}_\mu^\nu$.  The relevant components of the energy tensor $\mathbf{T}$ are easily computed from \refeq{Fcomp} and \refeq{starMcomp} to be:
 $$\bT_t^t  = \bT_r^r = \frac{-f}{A^2B^2} \left(\ell_x f + c \ell_y \frac{AB}{C^2}\right) - \ell,\qquad \bT_t^r = 0.$$
 We thus see that the solution satisfies the conditions of Eiesland's theorem, with $\Psi$ a constant.  We can take the components of the Killing field to be $k_0 = -\p_r C/(AB)$ and $k_1 = \p_t C/(AB)$.
  In that case $$ \bg(\bK,\bK) = - (\p_rC/B)^2 + (\p_t C/A)^2 = - g^{-1}(\nab C,\nab C),$$ and thus $\bK$ is timelike if and only if $\nab C$ is spacelike.  Since $d\bK$ is easily seen to be proportional to $dt\wedge dr$, the twist of $\bK$ vanishes: $\bK\wedge d\bK = 0$, and therefore $\bK$ is  hypersurface-orthogonal, i.e. the solution is static .  It would then be a matter of changing the coordinates $(t,r)$ into new coordinates $(t',r')$ that satisfy
 \beq\label{eq:coc} \bK t' = k_0 \frac{\p t'}{\p t} + k_1 \frac{\p t'}{\p r} = 1,\qquad \bK r'=k_0 \frac{\p r'}{\p t} + k_1 \frac{\p r'}{\p r} = 0,\eeq
 which is solvable by quadratures, in order for the line element of $\bg$ in the new coordinates to have the form
 $$ ds^2 = -P^2 dt'^2 +Q^2 dr'^2 + R^2 (d\theta^2 +\sin^2\theta d\phi^2),$$
 where $P,Q,R$ are functions of $r'$ only, and $R$ is not a constant. At this point we may again use that $\bG_0^0 - \bG_1^1 =  0$, where the indices now refer to the new coordinates $x^0 = t'$, $x^1 = r'$.  From the definition of $\bG_{\mu}^\nu$ we compute (see e.g. \cite{Eie25}) that
 $$ \bG_0^0 = \frac{1}{Q^3 R^2} \left( Q^3 + 2 Q' R R'-QR'^2-2QRR''\right),\quad
 \bG_1^1 = \frac{1}{PQ^2R^2}\left( PQ^2 - PR'^2 - 2 RR'P'\right).$$
 The equality of the two is now easily seen to imply that there exists a constant $\la$ such that $PQ = \la R'$.  Thus letting $\tau = \la t'$ and taking $(\tau,R,\theta,\phi)$ as the new coordinates, the metric takes the form
 $$ ds^2 = - \frac{P^2}{\la^2} d\tau^2 + \frac{\la^2}{P^2} dR^2 + R^2 (d\theta^2 + \sin^2 \theta d\phi^2),$$
 which is the form we assumed in Section 4 in order to derive the static spherically symmetric solution we found.

 \begin{rem} The above uniqueness result can most likely be strengthened by considering the remaining cases, of $\nab C$ being timelike, null, or zero, finding in each case what the solution reduces to, similar to the treatment of the cosmological vacuum solutions in \cite{SchWit10}.  We do not pursue this here, however.
 \end{rem}

\section{Precise Statement of the Main Result}
We are now in a position to give the precise version of the main result:
\begin{thm}\label{thm:formal}
Let $\ze:\RR^+\to \RR^+$ be any $C^2$ function satisfying the following conditions:
\begin{enumerate}
\item $\ze(\mu) = \mu + O(\mu^{5/4})$ as $\mu \to 0$.
\item $\ze'(\mu)>0$ and $\ze - \mu \ze'\geq 0$ for all $\mu>0$.
\item $\zeta'(\mu) + 2\mu\zeta''(\mu) \geq 0$ for all $\mu>0$.
\item $I_\zeta := \int_0^\infty \mu^{-7/4}\zeta(\mu) d\mu < \infty$.
\item There exists positive constants $\mu_0,J_\ze, K_\ze, L_\ze$ such that $ J_\ze \sqrt{\mu} - K_\ze \leq \ze(\mu) \leq J_\ze \sqrt{\mu}$, and $\frac{J_\ze}{2\mu^{1/2}} - \frac{L_\ze}{\mu} \leq \ze'(\mu) \leq  \frac{J_\ze}{2\mu^{1/2}}$ hold for $\mu>\mu_0$.
\end{enumerate}
    Let $\textsc{m}>0$ and $\textsc{q}\ne 0$ be two fixed real numbers, and let $\ep:=\textsc{m}/|\textsc{q}|$.  Define
    $$\ze_\be (\mu) := \frac{1}{\be^4} \ze(\be^4 \mu),\qquad\beta := \frac{ \textsc{q}^{3/2}I_\zeta}{\textsc{m}}.$$
    Let $\ell$ be any electromagnetic Lagrangian density function with the property that the electrostatic reduction of its Hamiltonian is the function $\ze_\be$.    Then the following hold:
\begin{enumerate}
\item The Einstein-Maxwell system (\ref{eq:Ein}-\ref{eq:Max}) with the electromagnetic Lagrangian density $\bL = - *\ell$ has a unique electrostatic, spherically symmetric, asymptotically flat solution, the maximal analytic extension of which is a Lorentzian manifold $(\M,\bg)$, called a {\em charged-particle spacetime.}  $\M$ is topologically equivalent to $\Rset^4$ minus a line.  There exists a global coordinate system $(t,r,\theta,\phi)$ on $\M$ with the property that $\bK:=\p_t$ is an everywhere-timelike Killing field, $r$ is the area-radius of rotation group orbits, $r=0$ is invariant under $\bK$, and $(\theta,\phi)$ are standard spherical coordinates on $\Sset^2$.
\item Any region in a spherically symmetric electrovacuum spacetime where the area-radius of the rotation group orbits has a spacelike gradient, is necessarily static as well, and in the case of no magnetic charge is isometric to a region in the corresponding charged-particle spacetime $(\M,\bg)$ defined above.
\item The line element of the metric $\bg$  has the form
    $$ds^2 = -e^{\xi} dt^2 +e^{-\xi} dr^2 + r^2 (d\theta^2 + \sin^2\theta d\phi^2),$$ where
    $$e^{\xi} := 1 - \frac{2m(r)}{r},\qquad m(r) := \int_0^r \zeta_\be\left(\frac{\textsc{q}^2}{2s^4}\right)s^2ds.$$
$m(r)$ is the {\em mass function} of the spacetime $\M$.  In particular, $m(0) = 0$, $m$ is increasing in $r$, and the ADM mass of $\M$ is $m(\infty) = \textsc{m}$. The mass function moreover has the following asymptotics:
    $$
    m(r) = \frac{A\epsilon^2}{2}r - \frac{B^2\ep^6}{2\textsc{m}^2} r^3 + o_2(r^3)\mbox{ as }r\to 0,\qquad
    m(r) = \textsc{m} - \frac{\textsc{q}^2}{2r} + O\left(\frac{1}{r^2}\right)\mbox{ as }r\to \infty,$$
where $A:= \frac{\sqrt{2}J_\ze}{I_\ze^2}$ and $B^2:=\frac{2K_\ze}{3 I_\ze^4}$. Consequently, $\bg_{rr}|_{r=0} = (1-A\ep^2)^{-1}>1$ and thus the metric has a conical singularity at $r=0$.
\item The mass function satisfies the following bounds
$$ \max\left\{ \frac{A\epsilon^2}{2}r - \frac{B^2\ep^6}{2\textsc{m}^2} r^3, \textsc{m} - \frac{\textsc{q}^2}{2r}\right\} \leq m(r) \leq \min\left\{ \frac{A\epsilon^2}{2}r,\textsc{m}\right\}.$$
Consequently, for $\ep$ small enough, $\bg_{tt} <0$ everywhere, which implies that there are no horizons in $\M$, hence the conical singularity at $r=0$ is naked.
\item $r=0$ is a curvature singularity, where the Kretschman scalar blows up like $r^{-2}$.  Any  spherically symmetric electrostatic spacetime homeomorphic to $\Rset^4$ minus a line, whose reduced Hamiltonian satisfies the first three conditions listed above for $\ze$ will necessarily have a singularity at $r=0$ which is at least this strong.
\item The spacetime $\M$ is endowed with an electromagnetic field $\bF = d\bA$, where $\bA = -\varphi(r) dt$,
$$\varphi  = \textsc{q} \int_r^\infty \zeta'_\be\left(\frac{\textsc{q}^2}{2s^4}\right) \frac{ds}{s^2}$$
is the electrostatic potential, and $\bE := e^{-\xi/2}d\varphi$ is the (flattened) {\em electric field}\footnote{I.e., $\bE = i_{\hat{\bK}}\bF$ where $\hat{\bK}$ is the {\em unit} vectorfield in the direction of $\bK$.}.  The potential $\varphi$ is smooth, monotone decreasing, and has the following asymptotics
$$ \sgn(\textsc{q})\varphi(r) = \frac{3\ep}{2}- \frac{A\ep^3}{2\textsc{m}} r+ O(r^3) \mbox{ as }r\to 0,\qquad \varphi(r) = \frac{\textsc{q}}{r} + O\left(\frac{1}{r^3}\right)\mbox{ as }r\to \infty.$$
Thus the total charge of the spacetime is $\textsc{q}$.  The electric field $\bE$ is smooth everywhere except at $r=0$ where it  has a point-defect, i.e. its magnitude has a finite limit but its direction is undefined.
\item The total electrostatic energy carried by a time slice is finite and is equal to $\textsc{m}$, therefore the mass of the spacetime is entirely of electric origin.
\item Radial null geodesics fall into the singularity and are thus incomplete in one direction, while null geodesics with  nonzero angular momentum are infinitely extendible in both directions. For $\ep$ small, these are deflected by the singularity by an amount that is proportional to $\ep^2$, and there are no trapped null geodesics.
\item Similarly, timelike geodesics and test-charge trajectories can only reach the singularity if they are radial, while non-radial ones have either bound orbits or escape orbits. The singularity at $r=0$, in contrast to the one in the superextremal RWN solution, is gravitationally attractive.
\end{enumerate}
\end{thm}
The only items in the above Theorem that we have not yet proved are the last two, regarding geodesics.  These will be established below:

\section{Geodesics and Test-Charge Trajectories}
Consider the following Lagrangian density, defined on a velocity bundle (see \cite{Chr99} for definitions):
$$L(\bq,\bv) = \half \bg_{\al\be}(\bq) \bv^\al \bv^\be + \fe \bA_\la(\bq)\bv^\la, $$
where $\bA$ is an electromagnetic vector potential defined on $\M$, $\bq\in\M$  and $\bv\in T_{\bq}\M$.  The corresponding action
$$ \A[\bq] = \int_\Rset L(\bq,\dot{\bq}) ds$$
is defined on curves $\bq(s) = (\bq^\al(s))$ in $\M$, where the dot represents differentiation with respect to an affine parameter $s$, which is related to arclength parametrization (``proper time" for timelike curves) $\tau$ by $\tau = \varmu s$.  The stationary points of this action are, for $\fe = 0$ and $\varmu^2 = 1,0,-1$ respectively timelike, null, and spacelike geodesics of the spacetime. Moreover, stationary points of the action with $\fe\ne0$ and $\fm^2=1$   represent possible worldlines of a ``test charge''\footnote{This is  a fairly standard treatment of test charges, cf. \cite{Dar59,Dar61}, \cite{GraBri60}, \cite{Car68} and \cite{Bon93}.  Notice however that it is not without conceptual problems:  If one thinks, e.g. as in \cite{Car68} and \cite{Bon93}, that the charged-particle spacetime in question is the spacetime of an electron, then one cannot possibly ``test'' an electron with an electron, i.e. a particle which has charge comparable to that of the charged-particle spacetime can by no stretch of imagination be considered a test particle, since its effect on the geometry of spacetime cannot be ignored, while on the other hand if the spacetime is to represent an elementary particle then of course no particle of smaller charge exists to test it with! The situation considered here is thus merely a cartoon, and proper treatment of this subject is postponed to a future paper, where we plan to consider charged-particle spacetimes featuring {\em two} point charges.} of mass $\varmu$ and charge $\fe$ in the spacetime $(\M,\bg)$ permeated by the electromagnetic field $\bF$.   The Euler-Lagrange equations for these geodesics are
$$\frac{Dd\bq^\al}{d\tau^2} = \frac{\fe}{\varmu} \bF^\al_\be(\bq)\frac{d\bq^\be}{d\tau},$$
where $D/d\tau$ is covariant differentiation operation on tangent vectors, and $\bF$ is the electromagnetic field tensor.

Let $$\bp_\al = \frac{\p L}{\p \bq^\al} = \bg_{\al\be}(\bq)\bv^\be + \fe \bA_\al(\bq)$$
be the canonical momenta, and $\Hc$ the corresponding Hamiltonian to $L$ defined by
$$ \Hc = \bp_\al \bv^\al - L = \half \bg^{\al\be}(\bq)(\bp_\al - \fe \bA_\al(\bq))(\bp_\be - \fe \bA_\be(\bq)) = \half |\bp|^2  - \fe \bA\cdot \bp + \frac{\fe^2}{2} |\bA|^2.$$
The geodesic equations in Hamiltonian form are
$\dot{\bq} = \frac{\p \Hc}{\p \bp}$ and $ \dot{\bp} = - \frac{\p \Hc}{\p\bq}$.
The first constant of motion is $\Hc$ itself, and from the normalization condition discussed above we have that along solutions $\Hc = -\half \varmu^2$.
Moreover, $\J(\bp,\bq)$ is a constant of motion iff $\{\J,\Hc\} = 0$.

Let us now take $\bg$ to be a point-charge metric, with empirical charge $\textsc{q}$ and mass $\textsc{m}$ (that is entirely of electric origin).  $\bg$ is  the spherically symmetric electrostatic solution to the Einstein-Maxwell system with a nonlinear aether law,  corresponding to a reduced Hamiltonian $\zeta_\be$ that is subject to the restrictions {\bf (R1-5)} discussed in the previous sections but otherwise arbitrary, and let $\bF$ be the corresponding electromagnetic field permeating this spacetime.  The line element of $\bg$ thus has the form
$$ ds_\bg^2 = -e^\xi dt^2 + e^{-\xi} dr^2 + r^2 (d\theta^2 +\sin^2\theta d\phi^2),$$
where $(\theta,\phi)$ are spherical coordinates on the standard unit sphere $\BS^2$. Moreover $\bF= d\bA$ where $\bA = \varphi(r)dt$, and the functions $\xi(r),\varphi(r)$ are given in terms of $\zeta$ by
\bna
e^\xi &=& 1 - \frac{2m(r)}{r};\qquad m(r) = \int_0^r \zeta_\be\left(\frac{\textsc{q}^2}{2s^4}\right)s^2ds;\\
\varphi & = &\textsc{q} \int_r^\infty \zeta'_\be(\frac{\textsc{q}^2}{2s^4}) \frac{ds}{s^2};\qquad \zeta_\be(x) = \be^{-4}\zeta(\be^4 x),
\ena
with $\be$ as in \refeq{valbe}.

With coordinates $(x^\al) = (t,r,\theta,\phi)$ thus chosen, we have
$$ p_t = -e^\xi \dot{t} +\fe \varphi;\quad p_r = e^{-\xi} \dot{r};\quad p_\theta = r^2\dot{\theta};\quad p_\phi = r^2\sin^2\theta \dot{\phi},$$
and
\beq\label{Ham}
\Hc = \half\left\{ -e^{-\xi}(p_t-\fe \varphi)^2 + e^{\xi} p_r^2 + \frac{1}{r^2} p_\theta^2 + \frac{1}{r^2 \sin^2\theta} p_\phi^2\right\}.
\eeq
Since $\p\Hc/\p t = \p\Hc/\p \phi = 0$ we get two more constants of motion, $p_t$ and  $p_\phi$, and we call the values that they take on a solution the {\em energy} $E$ and {\em angular momentum about the $\phi$ axis} $\Phi$:
$$ p_t \equiv E;\qquad p_\phi \equiv \Phi.$$
A fourth constant is also  found, by observing that
$$ \frac{d}{ds}r^2\dot{\theta} = \frac{d}{ds}p_\theta = - \frac{\p \Hc}{\p \theta} = \frac{2\Phi^2}{r^2\sin^3\theta}\cos\theta.$$
Thus, if one can arrange it so that $\theta=\pi/2$ when $\dot{\theta} = 0$, then $\ddot{\theta}=0$ as well, so that $\theta$ will stay constant at $\pi/2$.  But since we still have the freedom of choosing the axes for the two spherical coordinates, we can always choose the $\phi$ axis in such a way that the plane $\theta=\pi/2$ is the plane through the origin that contains the geodesic's initial velocity vector, so that $\dot{\theta} = 0$ initially.  We have thus shown that all geodesics of these particle-spacetimes are planar, and for any single geodesic we can always assume that it is contained in the equatorial plane $\theta = \pi/2$.

Now, evaluating \refeq{Ham} along a solution we obtain that
$
( -e^{-\xi}(E-\fe \varphi)^2 + e^{-\xi} \dot{r}^2 + \varmu^2 )r^2 \equiv -\Phi^2$,
which can be used to find an equation for $\dot{r}$:
$$
\dot{r}^2 = (E-\fe \varphi)^2 - \left(\frac{\Phi^2}{r^2} +\varmu^2\right) e^\xi.
$$
With four constants of motion the equations are reduced to a first-order system of ODEs.  We proceed to study this system, first for timelike and null geodesics ($\fe = 0$) and then for test-charges ($\fe \ne 0$).

\subsection{Geodesics}
For timelike (resp. null) geodesics, $\fe = 0$ and $\fm = 1$ (resp. $\fm = 0$). The equations are
\beq
\dot{t} = -e^{-\xi}E,\qquad
\dot{r} = \pm\sqrt{E^2 - \left(\frac{\Phi^2}{r^2}+\fm^2\right)e^\xi} \label{eq:rdot},\qquad
\dot{\phi} = \frac{1}{r^2}\Phi.
\eeq
We observe that for $\Phi=0$ the geodesic is radial, with $\dot{r} = \pm \sqrt{E^2 - \fm^2e^\xi}$. Thus, since $a^2\leq e^\xi<1$, for $E<\fm a$ there is no solution.  For $\fm a \leq E \leq \fm$  the geodesic does not have enough energy to escape, and will fall radially towards the singularity $r=0$ where curvature blows up. The geodesic reaches the singularity at a finite parameter value, since the function under the square root has only a simple zero.  This is also the case for geodesics with $E>\fm$ (and therefore for all null geodesics), the only difference being that they can reach infinity, and in fact have a well-defined asymptotic velocity: $(dr/dt)|_{r=\infty} = \sqrt{1 - \fm^2/E^2}$. All radial geodesics are therefore inextendible in one direction and thus incomplete.

Let $\Phi>0$ now. It is more convenient to reformulate the equations in terms of a reciprocal radial variable, and to measure length in units of $\textsc{m}$.  Let $$x := \frac{\textsc{m}}{r}.$$ From the $\dot{r}$ equation in \refeq{eq:rdot} one easily obtains
$$
\frac{d\phi}{dx} = \frac{\pm1}{\sqrt{\de^2 - (x^2 + \ga^2)(1-2xM(\frac{x}{\ep^2}))}},
$$
where
$$ \ga:= \frac{\textsc{m}\fm}{\Phi},\quad\de := \frac{\textsc{m}E}{\Phi},\qquad \ep:=\frac{\textsc{m}}{|\textsc{q}|},$$
and $M(y)$ is the {\em normalized mass function}
$$
M(y) := \frac{1}{2^{11/4}I_\zeta}\int_{I_\zeta^4 y^4/2}^\infty \zeta(\mu)\mu^{-7/4} d\mu = 1 -  \frac{1}{2^{11/4}I_\zeta}\int_{0}^{I_\zeta^4 y^4/2} \zeta(\mu)\mu^{-7/4} d\mu,$$
which, in view of \refeq{zetabounds}, satisfies the following bounds
$$ \begin{array}{rcl} \max\{1 - \half y , \frac{A}{2y} - \frac{B^2}{2y^3}\} \leq & M(y) & \leq \min\{ 1, \frac{A}{2y}\},\\ \max\{-\half,\frac{-A}{2y^2}\}\leq & M'(y) & \leq \min\{0,\frac{-A}{2y^2} + \frac{3B^2}{2y^4}\},\end{array}$$
where $$A:= \frac{\sqrt{2}J_\zeta}{I_\zeta^2} \geq 1,\qquad B^2:=\frac{2K_\zeta}{3I_\zeta^4},$$ are constants that depend only on the profile $\zeta$.  Note that for the RWN metric, $M_{RWN}(y) = 1-\half y$.
Let $$f_{\ga,\ep}(x):= (x^2 + \ga^2)\left(1-2xM\left(\frac{x}{\ep^2}\right)\right).$$
We note that for null geodesics, $\ga = 0$.  In that case, it is not hard to see that $f_{0,\ep}(x) = x^2e^{\xi(\textsc{m}/x)}$ is monotone increasing, for $\ep$ small enough, thus there is a unique $x_0>0$ such that $\de = f_{0,\ep}(x_0)$, and we must have $x<x_0$ along any null geodesic.  Hence $r_0 :=\textsc{m}/x_0$ is the {\em perihelion} for the geodesic, i.e. the closest it can get to the singularity. Furthermore, since $f_{0,\ep}$ has no critical points, it follows that there are no bounded orbits for null geodesics, so that all null geodesics with $\Phi>0$ can be extended in both directions to an infinite value for the parameter, and the events corresponding to those infinite values are points at infinity.  This can be easily seen from the $\dot{r}$ equation
$$ ds = \frac{\pm dr}{\sqrt{E^2-\frac{\Phi^2}{r^2}e^\xi}}\qquad \mbox{ for } r_0<r <\infty,$$
where $s$ is any affine parameter along the null geodesic. Since under the square-root has only a simple zero at $r=r_0$ and goes to a constant as $r \to \infty$ we see that the integral of the right-hand-side of the above is finite over any finite subinterval of $[r_0,\infty)$; and that this integral diverges only when the upper limit is infinite.

We thus have $(d\phi/dx)^2 = 1/(f_{0,\ep}(x_0)-f_{0,\ep}(x))$. For the Minkowski space, $f_{0,\ep}(x) = x^2$ and $x_0 = \delta$, thus the solution is $\phi(x) - \phi(0) = \sin^{-1}(x/\delta)$, which as expected is a straight line at a distance $1/\delta$ from the origin.  Since the particle-spacetime is always asymptotically Euclidean,  $\Phi/(\textsc{m}E) =1/\delta$ is the {\em impact parameter}, the distance from the central singularity of the initial asymptote of the geodesic. Geodesics with $1/\delta=0$ go straight to the central singularity, while geodesics with $1/\de>0$ will be deflected by it, as we will see below:

We find it more convenient to parameterize the family of null geodesics in the equatorial plane by their reciprocal perihelion $x_0$. Let $\phi_0$ denote the corresponding angle to $x_0$.  We then have
$$\phi - \phi_0 = \int_{x}^{x_0} \frac{\pm dx'}{\sqrt{ f_{0,\ep}(x_0)-f_{0,\ep}(x')}}.$$
The integral is convergent at the upper limit since as we said above  denominator has only a simple zero there.  Letting $\phi_\pm$ denote the two values for $\phi$  obtained at $x=0$, we have
$$\phi_+ - \phi_- = 2 \int_{0}^{x_0}  \frac {dx}{\sqrt{f_{0,\ep}(x_0)-f_{0,\ep}(x)}}=:\pi+d(x_0).$$
The quantity $d(x_0)$ represents the total deflexion of null geodesics due to spacetime curvature.  For the Minkowski space, one easily computes that
$ d(x_0) =  0,\forall x_0>0.$
By contrast for a static, spherically symmetric  spacetime whose line element is $-a^2dt^2 + (1/a^2) dr^2 + r^2 (d\theta^2 + \sin^2\theta d\phi^2)$ with $a^2<1$, so that it has a conical singularity at $r=0$, we have $f(x) = a^2 x^2$ and as a result $d(x_0) = \pi(\frac{1}{a} -1)>0$.  One can thus say that the null geodesics are ``bent" by the gravitational attraction of the conical singularity.   Below we will establish the lower bound $d(x_0)\geq c\ep^2$ when $x_0$ is sufficiently large, or equivalently, when $r_0/\textsc{m}$ is small. For large $r_0/\textsc{m}$ on the other hand, the geodesic stays far away from the singularity and thus its qualitative behavior is the same as in the RWN spacetime, as analyzed in \cite{GraBri60} and more extensively in \cite[\S40]{Cha83}.

As to the lower bound, we have, using that $a^2 x^2 \leq f_{0,\ep}(x) \leq a^2 x^2 +b^2$, with $a^2 = 1 - A \ep^2$ and $b^2 = 2B^2\ep^6$,
$$
\half(\pi+d(x_0)) \geq \int_0^{x_0} \frac{dx}{\sqrt{f(x_0) - a^2 x^2}} = \frac{1}{a} \sin^{-1} \frac{ax_0}{\sqrt{f(x_0)}} \geq \frac{1}{a} \sin^{-1} \frac{1}{\sqrt{1+ (b/ax_0)^2}}.$$

Let $0<c<A\pi/4$ be fixed and assume $x_0\geq \frac{B}{A\pi/4-c}\ep$.  We then have
\bea
\half(\pi+d(x_0)) &\geq& \frac{1}{\sqrt{1-A\ep^2}} \sin^{-1}\left(1 - \half\left(\frac{b}{ax_0}\right)^2 + O(\ep^4)\right) \\
& = &\left(1 + \frac{A}{2} \ep^2 + O(\ep^4)\right) \left(\frac{\pi}{2} - \frac{B}{x_0} \ep^3 + O(\ep^4)\right)\\
 &\geq & \frac{\pi}{2} + \left(\frac{A\pi}{4} - \frac{B\ep}{x_0}\right)\ep^2 + O(\ep^4),
\eea
 and thus $$d(x_0) \geq c \ep^2,$$
which is the desired lower bound.

Next we consider time-like geodesics with nonzero angular momentum.  One sees that $f_{\ga,\ep}(0) = \ga^2 > 0$, $f'_{\ga,\ep}(0) = -\ga^2<0$ and that the bounds for $M(y)$ obtained above imply that $f_{\ga,\ep}(x)$ grows like $x^2$ for large $x$.  Thus once again geodesics with $\de\geq\ga$ (i.e. those with $E\geq\fm$) will have a perihelion and are extendible to infinity in either direction.  In contrast to the null geodesics, however, there are bounded orbits for time-like geodesics.  This is because $f_{\ga,\ep}$ will  have at least one critical point.  Let $f_* := \min_x f_{\ga,\ep}(x)= f_{\ga,\ep}(x_*)$.  It then follows that there will be no solution with $\de <\sqrt{f_*}$, and that for $\sqrt{f_*} < \de<\ga$ the equation $f_{\ga,\ep}(x) = \de^2$ has at least two solutions, corresponding to the perihelion and the aphelion of a bound orbit.  There will also be at least one stable circular orbit, for $\de = \sqrt{f_*}$.
\subsection{Trajectories of test charges}
For the RWN metric, a detailed study of test-charge trajectories was done in \cite{Bon93}.  We are not aware of a comparable study done for the generalized RWN metrics with a nonlinear aether law, other than the particular case of the Born-Infeld  law \cite{Bre02} in the black hole regime, and some preliminary discussion of the general case in  \cite{DiaRub10}. Here we are going to analyze the qualitative behavior of all the test-charge trajectories in a given particle-spacetime with a small enough mass-to-charge ratio $\ep$.

Recall that the orbit of a single test particle of mass $\fm$ and charge $\fe$, in an electrostatic, spherically symmetric particle-spacetime of mass $\textsc{m}$ and charge $\textsc{q}$ satisfies the following system:
$$
\dot{t} =- e^{-\xi}(E-\fe \varphi),\qquad
\dot{r} = \pm\sqrt{(E-\fe\varphi)^2 - \left(\frac{\Phi^2}{r^2}+\fm^2\right)e^\xi},\qquad
\dot{\phi} = \frac{1}{r^2}\Phi.
$$
In addition, the orbit lies in a plane, which is taken to be the plane $\theta = \pi/2$.  The system is clearly invariant under the simultaneous sign change of $E$, $\fe$, $\Phi$ and the independent variable $s$.  Therefore, it is enough to consider the case $E\geq 0$.  Let $x:=\textsc{m}/{r}$ as before.  It then follows that
$$\dot{x} = \pm\frac{x^2}{\fm\textsc{m}}\sqrt{ \left(\ka - \rho N\left(\frac{x}{\ep^2}\right)\right)^2 - (1+\la^2x^2)\left(1-2xM\left(\frac{x}{\ep^2}\right)\right)},$$
where
$\ep= \frac{\textsc{m}}{|\textsc{q}|}$ as before and we have introduced new parameters$$\rho:= \frac{\fe/\fm}{\textsc{q}/\textsc{m}},\quad \ka := \frac{E}{\fm},\quad \la:=\frac{\Phi}{\fm\textsc{m}},$$
and where $N(y)$ is the {\em normalized electrostatic potential}:
$$ N(y) := \frac{1}{2^{7/4}I_\ze}\int_0^{I_\ze^4 y^4/2} \ze'(\mu) \mu^{-3/4} d\mu,$$
which in view of \refeq{zetabounds} satisfies the bound
$$
\max\left\{0, \frac{3}{2}-\frac{A}{2y}\right\} \leq N(y)  \leq \min\left\{y,\frac{3}{2} - \frac{A}{2y} + \frac{C}{3y^3} \right\}.
$$
Moreover, $N'(y) = \ze'(I_\ze^4 y^4/2)$, so that
$$
\max\left\{0,\frac{A}{y^2} - \frac{C}{y^4} \right\} \leq N'(y) \leq \min\left\{1,\frac{A}{y^2} \right\}.
$$
Note that for the RWN metric, $N_{RWN}(y) = y$.

$\ka$ and $\la$ are proportional to the energy and angular momentum of the test charge.  The ratio $\rho$ is positive in the case where the particle-spacetime and the test particle have charges of the same sign, and negative if they have the opposite sign. Setting $\rho = 0$ will reproduce the results obtained above for timelike geodesics. Our goal is to obtain a bifurcation diagram in the $\ka,\rho,\lambda$ parameter space for the above system, for a fixed small value of $\ep$.  It is clear that the diagram will be invariant under the simultaneous sign change of $\ka$ and $\rho$, and thus it is enough to restrict our attention to $\ka\geq 0$.

Let
$$g_{\rho,\ka,\ep}(x) := \ka- \rho N\left(\frac{x}{\ep^2}\right),\qquad h_{\la,\ep}:= (1+\la^2x^2)\left(1-2xM\left(\frac{x}{\ep^2}\right)\right).$$
Consider first the case of orbits with zero angular momentum: $\Phi = \la = 0$.  The motion of the test charge is then radial. We recall that $h_{0,\ep}(x) = e^{\xi(\textsc{m}/x)}$ Thus, $h_{0,\ep}$ is convex, decreasing,  $h_{0,\ep}(0) =1$,  $h'_{0,\ep}(0) = -2$ and $\lim_{x\to \infty} h_{0,\ep}(x) = a^2 =1-A\ep^2$.

On the other hand,  $g_{\rho,\ka,\ep}(0) = \ka$, $g'_{\rho,\ka,\ep}(0) = -\rho/\ep^2$ and that $\lim_{x\to\infty} g_{\rho,\ka,\ep}(x) = \ka - 3 \rho/2$.  Accordingly, there are three main parameter regimes to consider:

{\bf Case 1.} $\rho<0$. The two charges $\textsc{q}$ and $\fe$  thus have the opposite sign.  The function $g_{\rho,\ka,\ep}$ is increasing, and has a horizontal asymptote.  Thus $g_{\rho,\ka,\ep}^2$ must have the same properties, and in particular it will be concave and increasing, while we have already established that $h_{0,\ep}$  is  convex and decreasing.  It then follows that  there will be no trajectory with $\ka\leq a+3\rho/2$, for
$a + 3\rho/2<\ka<1$ the trajectory is bounded, and corresponds to a test charge that cannot escape and falls radially inward and into the singularity, and for $\ka\geq 1$ the test charge has enough energy to escape to infinity.

{\bf Case 2.} $\rho>0$, $\ka>3\rho/2$. The two charges thus have the same sign, and the test charge has relatively high energy.  The function $g_{\rho,\ka,\ep}$ is decreasing, has a positive asymptotic value, and is convex. Thus $g^2_{\rho,\ka,\ep}$ will have these same three properties as well.  In order to find the number of intersections of the graphs of $g^2_{\rho,\ka,\ep}$ and $h_{0,\ep}$, one needs to compare the values of the derivatives of these two functions at $x=0$ and near $x=\infty$.  we have
$$ 2g_{\rho,\ka,\ep}(0) g'_{\rho,\ka,\ep}(0) - h'_{0,\ep}(0) = 1-\frac{\ka\rho}{\ep^2},$$
while, using the large $y$ asymptotics established above for the two functions $M(y)$ and $N(y)$, we have
$$ 2g_{\rho,\ka,\ep}(x) g'_{\rho,\ka,\ep}(x) - h'_{0,\ep}(x) = -\rho\ep^2\left(\ka - \frac{3}{2} \rho\right) \frac{1}{x^2} + O\left(\frac{1}{x^3}\right) < 0 \quad\mbox{ as }x\to\infty.$$
Thus we need to distinguish two subcases:

{\bf(2a)} $\ka < \ep^2/\rho= \frac{\fm \textsc{m}}{\fe \textsc{q}}$. (Note that this requires $\rho^2<2\ep^2/3$, or equivalently, $\fe/\fm < \sqrt{2/3}$, which does not allow the test charge to have a charge-to-mass ratio typical of actual elementary particles.) In this case there will be a critical value of $\ka$, say $\ka_c\in (3\rho/2,a+3\rho/2)$, such that the graphs of the two functions $g^2_{\rho,\ka_c,\ep}$ and $h_{0,\ep}$ are tangent to each other at a positive $x=x_c$.  Thus for $\ka<\ka_c$ there will be no trajectories,  for $\ka = \ka_c$ there will be a stable equilibrium, corresponding to the test charge remaining at rest with respect to the singularity, for $\ka_c<k<a+3\rho/2$ the test charge shuttles back and forth between a perihelion and an aphelion, which grow further apart as $\ka$ is increased, for  $a+3\rho/2 \leq \ka< 1$ the perihelion coincides with the singularity, and for $\ka\geq 1$ the aphelion is at $r=\infty$, i.e. the test charge is allowed to escape.

{\bf(2b)}$\ka\geq \ep^2/\rho$: The two above-mentioned graphs can have no point of tangency, irrespective of $\ka$. Moreover, there is no intersection at a positive $x$, and hence no trajectory, with $\ka \leq1$ (i.e. for $E\leq\fm$).  For $1<\ka<a+3\rho/2$ there is a single intersection, which corresponds to the perihelion of a hyperbolic trajectory that avoids the singularity, and for $\ka>a+3\rho/2$ the test charge in one direction falls into the singularity and in the other direction is allowed to escape to infinity.

{\bf Case 3}. $\rho>0$ and $ \ka<3\rho/2$.  The charges have the same sign and the test charge has relatively low energy. The function $g_{\rho,\ka,\ep}$ has a negative asymptotic value, thus the graph of $g^2_{\rho,\ka,\ep}$ will be tangent to the $x$-axis at some  point $x=x_*$, which will be where the global minimum is achieved, and is asymptotic to $(\ka-3\rho/2)^2$. There are no trajectories with $3\rho/2-a<\ka<1$. For $\ka< \min\{1,3\rho/2-a\}$ the  trajectories correspond to falling-in test particles.  For each $1<\ka<3\rho/2-a$ there is an avoiding trajectory as well as one that falls in.  For $3\rho/2-a<\ka<3\rho/2$ all trajectories avoid the singularity.

\begin{center}
\begin{figure}[h]
\hspace{1.5in}
\includegraphics[scale=0.4]{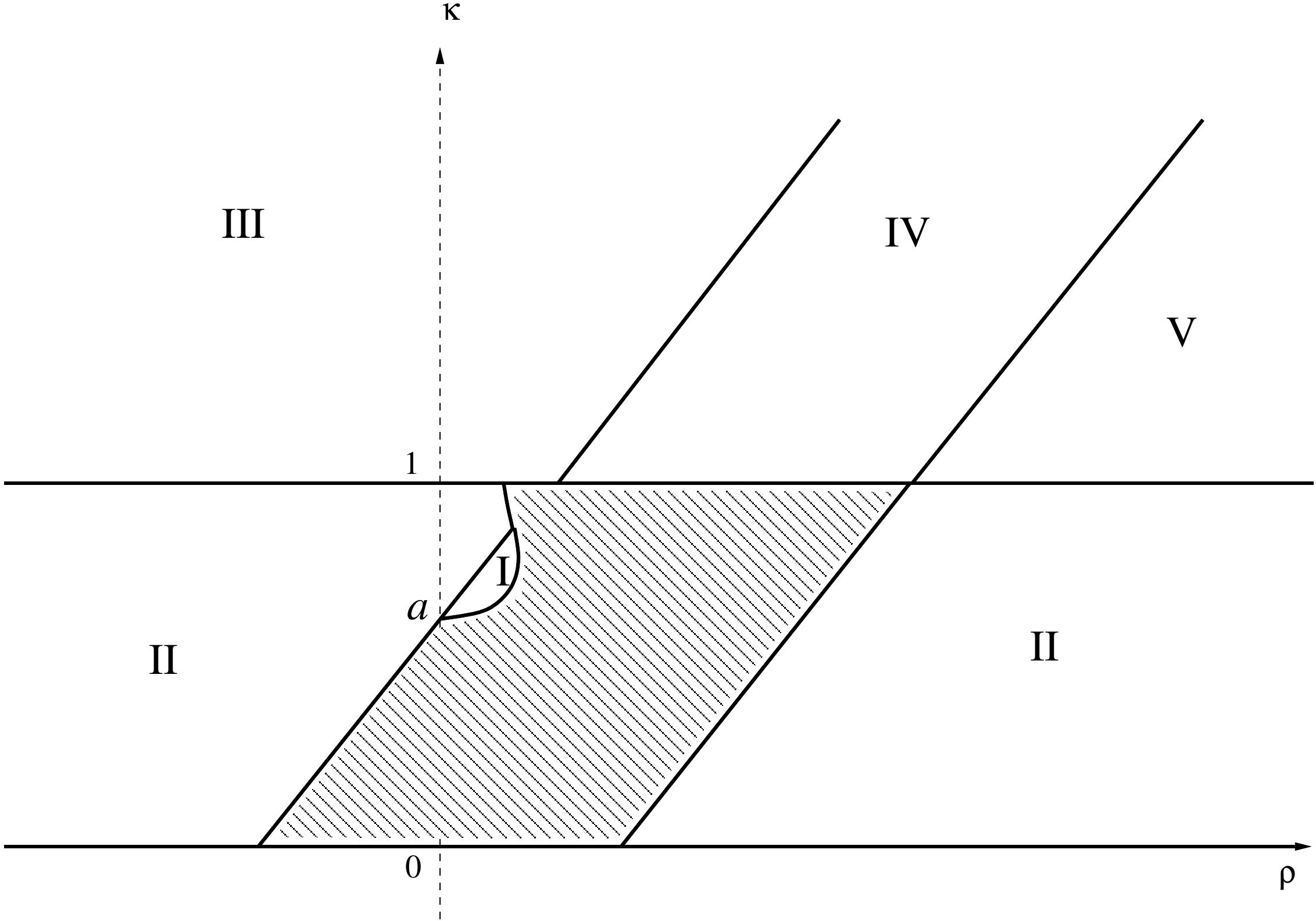}
\caption{\label{fig:zam}Bifurcation diagram for trajectories with no angular momentum}
\end{figure}
\end{center}

\vspace{-35pt}
Based on the above analysis, the $\ka$-$\rho$ parameter plane is divided into several regions as in Figure~\ref{fig:zam}.
The shaded region in this figure is a forbidden zone: there are no trajectories for $(\ka,\rho)$ in this region.  Region {\bf I} corresponds to shuttling orbits, region {\bf II} to test particles that are trapped and fall into the singularity.  Particles with parameters in region {\bf III} can also fall into the center, but have enough energy so that in the opposite direction they are allowed to escape to infinity.  Region {\bf IV} corresponds to hyperbolic trajectories that avoid the singularity altogether, while each point  in region {\bf V} corresponds to two trajectories, one that falls into the center and another that avoids it.

\begin{center}
\begin{figure}[h]
\hspace{1.5in}
\includegraphics[scale=0.4]{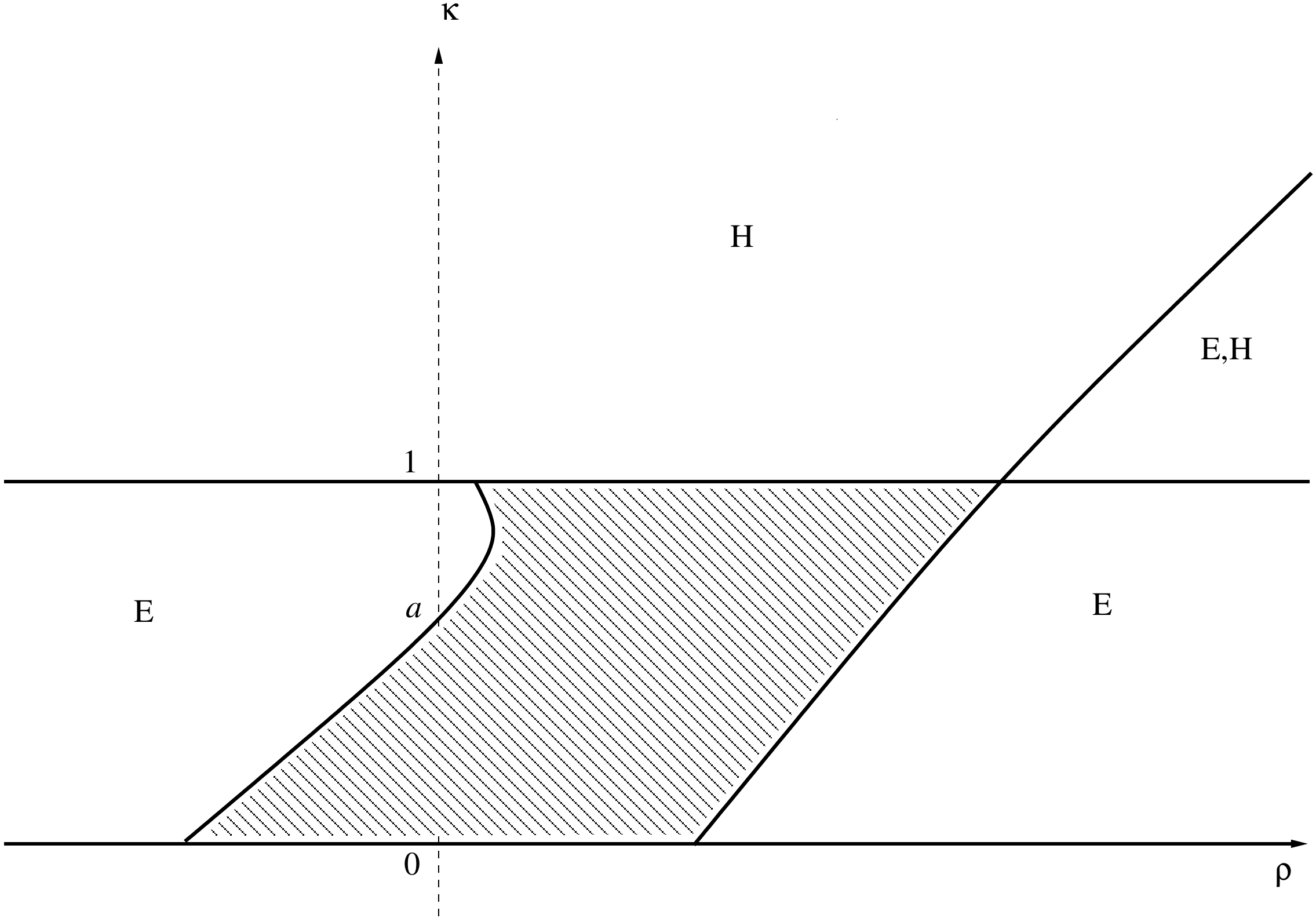}
\caption{\label{fig:zam2}Bifurcation diagram for trajectories with positive angular momentum}
\end{figure}
\end{center}

Next, we consider charged trajectories with nonzero angular momentum, i.e. $\la>0$.  The only thing that changes in the above analysis is the behavior of $h_{\la,\ep}$, which now grows like $x^2$ for large $x$, and has a global minimum at a positive $x$.  As a result, there will be no trajectories falling into the singularity.  The trajectories in fact will be similar to classical Keplerian trajectories, divided into {\em bound} and {\em escape} (scattering) orbits. The following  diagram (Fig.~\ref{fig:zam2}) shows the various parameter regimes for each type of orbit.
\vspace{-22pt}
The shaded region in Fig.~\ref{fig:zam2} is a forbidden one.  The parameters in regions labeled $E$ give rise to bound orbits, while those in regions labeled $H$ give rise to escape (scattering) ones. One observes that, in sharp contrast to the case of superextremal RWN studied in \cite{Bon93}, there is no evidence here of gravity being repulsive in the vicinity of the naked conical singularity at the center of these particle-spacetimes. On the contrary, the singularity appears to be attractive.

\bigskip

\section*{Acknowledgements:} I was introduced to nonlinear electrodynamics by my dear friend and colleague Michael Kiessling, and I also owe him the initial impetus for studying solutions with no horizon.  I have benefitted greatly from his help and encouragement throughout this project, and I am indebted to him for his critical reading of many drafts.  I would also like to thank the anonymous referee for constructive comments, and the Institute for Advanced Study for their hospitality while I was a member there during the final stage of this work.

{\small
\baselineskip=13pt

}

\begin{thebibliography}{}

\bibitem{Rei16}
Reissner, H.
{\"Uber die Eigengravitation des elektrischen Feldes nach der Einsteinschen Theorie.}
{\em Ann. Phys.} (Germany) {\bf 50}, 106-120 (1916).

\bibitem{Nor18}
Nordstr\"om, G.
{On the energy of the gravitational field in Einstein's theory},
{\em Proc. Kon. Ned. Akad. Wet.} {\bf 20}, 1238--1245 (1918).

\bibitem{Wey17}
Weyl, H.
{Zur Gravitationstheorie.}
{\em Ann. Phys.} (Germany) {\bf 54}, 117 (1917).

\bibitem{MisThoWhe79}
Misner, C. W., Thorne, K. S., and Wheeler, J. A.,
\textit{Gravitation},
W. H. Freeman \& Co, New York, 1973.

\bibitem{ADM61}
R. Arnowitt, S. Deser, and C. W. Misner,
{Coordinate invariance and energy expressions in
general relativity,}
{\em Phys. Rev.} {\bf 122},  997--1006 (1961).

\bibitem{HawEll73}
Hawking, S., and Ellis, G.,
\textit{The Large Scale Structure of Space-Time}.
Cambridge Univ. Press, Cambridge, 1973.

\bibitem{KraSteHerMac80}
Stephani, H., Kramer, D., MacCallum, M., Hoenselaers, C., and Herlt, E.,
\textit{Exact Solutions of Einstein's Field Equations},
Cambridge University Press, Cambridge 2003.

\bibitem{Wei39}
Weisskopf, V. F.,
{On the self-energy and the electromagnetic field of the electron},
{\em Phys. Rev.} {\bf 56}, 72--85 (1939).

\bibitem{Fey64}
Feynman, R. P.,
\textit{Lectures in Physics,} Vol. 2, Chap. 28, Addison-Wesley, Reading, Mass. (1964).

\bibitem{Wei95}
Weinberg, S.,
\textit{The Quantum Theory of Fields}, Vol. I, Cambridge University Press, Cambridge (1995), p.31.

\bibitem{ADM60}
 Arnowitt, R., Deser, S., and Misner, C. W.,
{Gravitational-electromagnetic coupling and the classical self-energy problem},
 {\em Phys. Rev.} {\bf 120}, 313--319 (1960).

\bibitem{ApKi01}
Appel, W. and Kiessling, M. K.-H.,
{Mass and spin renormalization in Lorentz electrodynamics},
{\em Ann. Phys.} (N.Y.) \textbf{289}, 24--83 (2001).

\bibitem{Kie10}
Kiessling, M. K.-H.,
{\em personal communication.}

\bibitem{Bor33}
 Born, M.,
{Modified field equations with a finite radius of the electron},
 {\em Nature} {\bf 132}, 1004 (1933).

\bibitem{Mie13}
 Mie, G.,
{Grundlagen einer Theorie der Materie},
 {\em Ann. Phys.} {\bf 37}, 511--534 (1912); {\em ibid.} {\bf 39}, 1--40 (1912); {\em ibid.} {\bf 40}, 1--66 (1913).

\bibitem{Sch03}
Schwarzschild, K.
{Zur Elektrodynamik, I. Zwei Formen des Princips der Action in der Elektronentheorie},
{\em Nachrichten von der Gesellschaft der Wissenschaften  zu Goettingen,} 126-131 (1903).

\bibitem{BorInf33}
 Born, M. and Infeld, L.
 {Foundation of the new field theory},
 {\em Nature} {\bf 132}, 1004 (1933); {\em Proc. Roy. Soc. London} A {\bf 144}, 425--451 (1934).

\bibitem{Kie04}
 Kiessling, M. H.-K.,
 {Electromagnetic field theory without divergence problems 1. The Born legacy},
 {\em J. Stat. Phys.} {\bf 116}, 1057--1120 (2004).

\bibitem{Kie11}
Kiessling, M. H.-K.,
{On the motion of point defects in relativistic fields},
{\em preprint}, (2011).

\bibitem{Hof35b}
Hoffmann, B.,
{Gravitational and electromagnetic mass in the Born-Infeld electrodynamics},
{\em Phys. Rev.} {\bf 47}, 877-880 (1935).

 \bibitem{InfHof37}
 Hoffmann, B., and Infeld, L.,
{On the choice of the action function in the new field theory},
 {\em Phys. Rev.} {\bf 51}, 765--773 (1937).

 \bibitem{Rao37}
 Madhava Rao, B. S.,
 {Generalized action-functions in Born's electro-dynamics},
 {\em Proc. Indian Acad. Sci., Sec.} A {\bf 6}, 158--173 (1937).

 \bibitem{PelTor69}
 Pellicer, R., and Torrence, R. J.,
 {Nonlinear electrodynamics and general relativity},
 {\em J. Math. Phys.} {\bf 10}, 1718--1723 (1969).

\bibitem{Dem86}
Demianski, M.,
{Static electromagnetic geon},
{\em Foundations of Physics} {\bf 16}, (1986).

\bibitem{AyoGar98}
Ay\'on-Beato, E., and Garc\'\i a, A.,
{Regular black hole in general relativity coupled to nonlinear electrodynamics},
{\em Phys. Rev. Lett.} {\bf 80}, 5056--5059 (1998).

\bibitem{Bro01}
 Bronnikov, K. A.,
 {Regular magnetic black holes and monopoles from nonlinear electrodynamics},
 {\em Phys. Rev.} {\bf D 63}, 044005 (2001).

  \bibitem{Dym04}
 Dymnikova, I.,
 {Regular electrically charged vacuum structures with de Sitter centre in nonlinear electrodynamics coupled to general relativity},
{\em Classical Quantum Gravity} {\bf21}, 4417 (2004).

\bibitem{Cir05JMP}
 Cirilo-Lombardo, D. J.,
 {New spherically symmetric monopole and regular solutions in Einstein-Born-Infeld theories},
 {\em J. Math. Phys.} {\bf 46}, 042501 (2005);
 {Rotating charged black holes in Einstein-Born-Infeld theories and their ADM mass},
 {\em Gen. Relativ. Gravit.} {\bf 37}, 847--856 (2005).


 \bibitem{BroMelShiSta79}
 Bronnikov, K. A., Melnikov, V. N., Shikin, G. N., and Staniukovich, K. P.,
 {Scalar, electromagnetic, and gravitational fields interaction: Particlelike solutions},
 {\em Ann.  Phys.} (N.Y.){\bf 118}, 84 (1979).

\bibitem{Bro00}
Bronnikov, K. A.,
{ Comment on ``Regular Black Hole in General Relativity Coupled to Nonlinear Electrodynamics",}
{\em Phys. Rev. Lett.} {\bf 85}, 4641 (2000).

\bibitem{Bir23}
Birkhoff, G. D.,
{\em Relativity and Modern Physics,}
Harvard University Press, Cambridge MA (1923), p. 253.

\bibitem{Jeb21}
Jebsen, J. T.,
{\"Uber die allgemeinen kugelsymmetrischen L\"osungen der Einsteinschen Gravitationsgleichungen im Vakuum,}
{\em Arkiv f\"or Matematik, Astronomi och Fysik} {\bf 15}, 1-9 (1921);  English translation in: {\em Gen. Relativ. Gravit.} {\bf 37}, 2253--2259  (2005).

\bibitem{EhlKra06}
Ehlers, J. and Krasi\'nski, A.,
{Comment on the paper by J. T. Jebsen reprinted in Gen. Rel. Grav. 37, 2253--2259 (2005)},
{\em Gen. Relativ. Gravit.} {\bf 38}, 1329--1330 (2006).

\bibitem{Eie25}
Eiesland, J. A.,
{The group of motions of an Einstein space,}
{\em Trans. Am. Math. Soc.} {\bf 27}, 213--245 (1925).

\bibitem{Eie21}
Eiesland, J. A.,
{\em Bull. Am. Math. Soc.} {\bf 27}, 410 (1921).

\bibitem{Hof32}
Hoffmann, B.,
{On the spherically symmetric field in relativity},
{\em Quart. J. Math.} {\bf 3}, 226--237 (1932).

\bibitem{SchWit10}
Schleich, K., and Witt, D. M.,
{A simple proof of Birkhoff's theorem for cosmological constant},
{\em J. Math. Phys.} {\bf 51}, 112502 (2010).

\bibitem{Chr99}
 Christodoulou, D.,
 \textit{The Action Principle and Partial Differential Equations}, Chap. 6,
 Princeton University Press, Princeton NJ (1999).

\bibitem{Ple68}
Plebanski, J.,
\textit{Lectures on Nonlinear Electrodynamics},
NORDITA, Copenhagen (1968).

\bibitem{Tah10}
Tahvildar-Zadeh, A. S.,
{One- and two-Killing field reductions of the Einstein-Maxwell system with arbitrary constitutive laws},
{\em in preparation.}

 \bibitem{Bia83}
Bia\l inicki-Birula, I.,
{Nonlinear electrodynamics: variations on a theme by Born and Infeld},
in {\em Quantum Theory of Particles and Fields}, B. Jancewicz and J. Lukierski, eds.,
 World Scientific, Singapore (1983), pp. 31--48.

 \bibitem{Hof35}
 Hoffmann, B.,
 {On the new field theory},
 {\em Proc. Roy. Soc. London} A {\bf 148}, 353--364 (1935).

\bibitem{Dar59}
Darwin, C.,
{The gravity field of a particle},
{\em Proc. Roy. Soc. London.} A {\bf 249}, 180--194 (1959).

\bibitem{Dar61}
Darwin, C.,
{The gravity field of a particle, II},
{\em Proc. Roy. Soc. London} A {\bf 263}, 39--50 (1961).

\bibitem{GraBri60}
Graves, J. C. and Brill, D. R.,
{Oscillatory character of Reissner-N\"ordstrom metric for an ideal charged wormhole},
{\em Phys. Rev.} {\bf 120}, 1507--1513 (1960).



\bibitem{Car68}
Carter, B.
{Global structure of the Kerr family of gravitational fields},
{\em Phys. Rev.} {\bf 174}, 1559--157 (1968).

\bibitem{Bon93}
Bonnor, W. B.,
{The equilibrium of a charged test particle in the field of a spherical charged mass in general
relativity},
{\em Classical \& Quantum Gravity} {\bf 10}, 2077--2082 (1993).

\bibitem{Cha83}
Chandrasekhar, S.,
\textit{The Mathematical Theory of Black Holes},
Oxford University Press, New York (1983).

\bibitem{Bre02}
Bret\'on, N.,
{Geodesic structure of the Born-Infeld black hole,} {\em Class. and Quantum Grav.} {\bf 19}, 601--612 (2002).

\bibitem{DiaRub10}
Diaz-Alonso, J. and Rubiera-Garcia, D.,
{Electrostatic spherically symmetric configurations in gravitating nonlinear-electrodynamics},
{Phys. Rev.}  {\bf D 81}, 064021 (2010).



\end{thebibliography}
\end{document}